\documentclass[12pt]{iopart}
\usepackage{bm}
\usepackage{hyperref}
\usepackage{framed}
\usepackage{subfigure}
\usepackage{hyperref}   
\usepackage{graphicx}
\usepackage{iopams}
\expandafter\let\csname equation*\endcsname\relax 
\expandafter\let\csname endequation*\endcsname\relax 
\usepackage{amsmath}
\usepackage{color}
\usepackage{yfonts}

\usepackage{hyperref}
\hypersetup{
    colorlinks=true,
    linkcolor=blue,
    filecolor=magenta,
    urlcolor=cyan,
}
\newcommand{\OU}{\mathrm{R}}
\newcommand{\IN}{\mathrm{L}}

\newcommand{\s}{l}
\newcommand{\tp}{{tp}}
\newcommand{\lr}{{lr}}
\newcommand{\ex}{{ex}}
\newcommand{\cst}{\mathrm{const}}
\newcommand{\eq}{{eq}}

\begin{document}

\title{Rotational superradiance with Bogoliubov dispersion}

\author{Sam Patrick}
\address{School of Mathematical Sciences, University of Nottingham, Nottingham, NG7 2FD, United Kingdom}

\date{\today}

\begin{abstract}
Rotational superradiance affects the dynamics of many rotating systems in nature, through either stimulated or spontaneous extraction of energy and angular momentum.
By now, this process is well-studied in the relativistic setting, where systems are intrinsically dispersion-free.
In many condensed matter systems, however, dispersion is an unavoidable aspect of the description for the short wavelength modes.
For these systems, how might one expect superradiance to be modified?
In this work, an answer to this question is provided using an illustrative example.
The scattering of linear excitations of a Bose-Einstein condensate are studied in the presence of a rotating, draining vortex flow using the full Bogoliubov dispersion relation.
It is shown that dispersion suppresses the extraction of energy and angular momentum, firstly, by decreasing the superradiant bandwidth, and secondly, by preventing high-angular momentum modes from superradiating.
\end{abstract}

\maketitle


\section{Introduction}

Rotational superradiance is a fundamental process involving the amplification of waves, with implications for the dynamics of many rotating systems in nature.
In general relativity, it allows for the extraction of energy and angular momentum from rotating black holes, which will eventually shed almost all of their angular momentum through sponteneous emission \cite{page1976particle}.
Recently, it has been demonstrated that this kind of amplification also occurs in condensed matter systems \cite{torres2017rotational}.
One of the fundamental differences with relativity is that the atomic nature of condensed matter systems gives rise to non-linear dispersion for short wavelengths.
The aim of this work will be to show how superradiance is altered by a quartic (Bogoliubov) modification to the dispersion relation.

In general, the term superradiance is used to describe the energy enhancement of radiation, which allows incident waves to extract energy from the system they scatter with (see \cite{bekenstein1998many,brito2015superradiance} for a review).
It is perhaps most famously associated with energy extraction from rotating black holes, and has played a central role in fashioning modern understanding of these elusive cosmic entities \cite{misner1972stability,starobinsky1974waves,starobinsky1974electro}.
Superradiance is sometimes called the wave equivalent of the Penrose process~\cite{penrose1971extraction}, where a black hole loses mass and angular momentum by absorbing particles with negative energy.
Indeed, it was along this last line of enquiry that the principles of black hole thermodynamics were established~\cite{bekenstein1994entropy}, leading ultimately to Hawking's discovery of black hole evaporation \cite{hawking1974explosions}.

Since Unruh's demonstration that certain fluids have the capacity to mimick features of spacetime \cite{unruh1981experimental} (which subsequently  developed into the field of analogue gravity \cite{barcelo2011analogue}),
there has been a surge of interest in understanding superradiance around fluid flows (see e.g. \cite{basak2003superresonance,basak2003reflection,richartz2015rotating}).
The prototypical example of a superradiating fluid system is a rotating draining vortex flow.
In the uniform density approximation, the most general solution of the irrotational and incompressiblity conditions gives the draining bathtub (DBT) model for the vortex \cite{dolan2011AB,dolan2013scattering}.
This system shares many features with Kerr black holes, in particular, the existence of a horizon and an ergoregion \cite{dolan2012resonances}.

The study of such analogue systems is well-motivated from several angles. 
Firstly, they can be used to test gravitational phenomena under controlled laboratory settings.
This is important since direct detection of many interesting processes (in particular, astrophysical Hawking radiation and superradiance) is beyond the scope of existing technology.
Secondly, analogue systems can probe how certain phenomena behave under modifications to the governing theory.
For example, it was demonstrated in \cite{unruh1995sonic,corley1996spectrum} using modified dispersion relations that the Hawking effect does not rely on the high energy behaviour of the theory, an important realisation given the lack of knowledge concerning physics below the Planck scale.
Finally, analogue systems are perfectly poised to mediate the transfer of techniques and ideas between different fields of physics.
This is exemplified by recent findings that the relaxation of draining vortices can be understood using the physics of light-rings \cite{torres2020quasinormal}, a concept routinely applied in relativity \cite{cardoso2009geodesic}.
Whilst analogue gravity is usually geared toward providing new input into gravitational physics, this demonstrates that analogue gravity really is a two way street.

To date, experimental efforts to detect superradiance have focussed on classical fluids.
The first direct detection of rotational superradiance was performed using surface waves in a water tank experiment containing a draining vortex \cite{torres2017rotational}.
Although the analogy to black hole physics using surface waves is mathematically precise only in shallow water, and most modelling efforts at the time had dealt solely with this regime, the amplification was in fact measured closer to the deep water regime where the system is strongly dispersive.
Following on from this, a theoretical basis for superradiance in dispersive systems was outlined in \cite{patrick2020superradiance}, and it was shown that amplification is also expected in deep water.
Recently, proposals to observe superradiance of acoustic beams from rotating absorbers were outlined in \cite{faccio2019superradiant2,gooding2018superradiant}, and the first experimental results were obtained in \cite{cromb2020amplification}.

Another promising analogue gravity system involves the linear excitations of a Bose-Einstein condensate (BEC) \cite{garay2000sonic,barcelo2001analogue}.
Indeed, a series of experiments on one dimensional BECs have successfully tested various aspects of the Hawking effect \cite{steinhauer2016observation,denova2019observation,kolobov2019spontaneous}.
The existence of vortices in BECs also raises the possibility of observing superradiance there.
After Pitaevskii's original treatment of the line vortex in a Bose gas \cite{pitaevskii1961vortex}, vortices in BECs have received much attention in the literature (see, e.g. \cite{fetter2001vortices} for a review).
Two important distinctions with the classical case are that, in a BEC, the fluid is truly irrotational (in classical systems this is only an approximation) and the circulation of the flow is quantised.
This opens up the tantalising possibility of studying gravitational phenomena in an effective spacetime which displays quantum behaviour.
It is tempting to go further and say that this line of research may even provide insight into the quantum nature of gravity.

DBT superradiance in a BEC has been studied in \cite{basak2005analog,federici2006superradiance,ghazanfari2014acoustic,demirkaya2019analog,demirkaya2020acoustic} and the potential difficulty in experimentally realising draining flows has led some authors to consider superradiance in purely rotating systems \cite{slatyer2005superradiant,giacomelli2020ergoregion}.
With exception of \cite{giacomelli2020ergoregion,faccio2019superradiant} which dealt with non-draining flows, the effects of short wavelength dispersion are usually ignored on the grounds that the quantum pressure term (which gives rise to the modified dispersion) remains small on suitably large length scales.
Within this approximation, often called the hydrodynamic approximation, the equations of motion become second order in spatial derivatives, which means that one benefits from all the standard techniques used to study superradiance based on the properties of second order differential equations.
The problem with this approximation is that systems containing horizons naturally probe the short-wavelength behaviour of the system, since out-going waves become increasingly blue-shifted as they are traced back toward their place of origin.
Hence, a fully consistent treatment requires the inclusion of dispersion.
It will be the aim of the coming sections to develop such a treatment.

\section{Objectives}

The main tool used in this work will be a combination of multiple scale analysis with matched asymptotic techniques.
These methods, often brought under the umbrella term of WKB approximations \cite{buhler2014waves}, provide an intuitive picture of wave scattering by recasting the problem in terms of effective particles, with accuracy improving in the limit of short wavelengths.
This approach has been shown to yield accurate predictions for high angular momentum modes when compared with laboratory experiments \cite{torres2020quasinormal}.
In fact, the method also provides a good indication of the general behaviour of the scattering coefficients even for low angular momentum modes \cite{patrick2020superradiance}.

There are several questions this paper will aim to address.
Firstly, the well-known condition for black hole superradiance is,
\begin{equation} \label{sr_bh}
\omega<m\Omega_h^\mathrm{rot},
\end{equation}
where $\omega$ and $m$ determine the wave energy and angular momentum respectively, and $\Omega_h^\mathrm{rot}$ is the rotational velocity of the spacetime on the horizon in radians per second.
The same condition applies in fluid systems in the non-dispersive approximation, with $\Omega_h^\mathrm{rot}$ now playing the role of the rotational fluid velocity.
It was shown in \cite{patrick2020superradiance} that a similar condition holds for the amplification of deep water gravity waves around the DBT.
The first task here will be to establish equivalent of \eqref{sr_bh} for the Bogoliubov dispersion relation.

The second aim will be to investigate the behaviour of the reflection coefficient, following the method established in \cite{patrick2020superradiance}.
To do this, the solutions of the dispersion relation will be classified into separated categories represented by a ``Feynman'' diagram, which depict the mode trajectories through phase space. 
Each category corresponds to a particular structure in the transfer matrix, which is used to compute the scattering coefficients.
This will allow for a comparison of the amount of amplification between the dispersive and non-dispersive cases.

Finally, the system will be quantized following the procedure outline in \cite{unruh1974second} and the spontaneous radiation of energy and angular momentum due to amplification of vacuum fluctuations studied.
Note that since only scattering in the presence of real turning points (defined later) of the dispersion relation is considered, the rates of energy and angular momentum loss include only the contributions of superradiant modes.
In reality, there will also be a (nearly) thermal spectrum of modes emitted by the vortex, resulting from the additional scattering which is not captured by the turning points.
This thermal emission is simply the Hawking effect, which has been well-studied in the context of analogue models of gravity, see e.g. \cite{coutant2012black,coutant2016imprint}, and could feasibly be incorporated into the present framework using the more thorough scattering treatments described therein.

The remainder of this paper is structured as follows.
In Section~\ref{sec:setup}, the equations governing the background and excitations of a BEC are introduced.
In Section~\ref{sec:wkb}, the WKB method is used to find approximate solutions to the wave equation, and it is shown that scattering between different WKB modes can be estimated by computing the amount of mode mixing around turning points.
These techniques are then applied in Sections~\ref{sec:nondisp} and \ref{sec:disp} to characterise the different scattering possibilities in the non-dispersive and dispersive cases respectively.
In Section~\ref{sec:spont}, the spontaneous emission of superradiant modes due to vacuum fluctuations is computed, and the rates of energy and angular momentum loss from the vortex compared for the non-dispersive and dispersive cases.
Finally, the relevance of these findings are discussed in Section~\ref{sec:discuss}.

\section{Set-up} \label{sec:setup}

In a BEC, the mean-field condensate wavefunction $\Psi(t,\mathbf{x})$ satisfies the Gross-Pitaevskii equation (GPE),
\begin{equation}
i\hbar\partial_t\Psi = -\frac{\hbar^2}{2M}\nabla^2\Psi + V(\mathbf{x})\Psi + g|\Psi|^2\Psi = 0,
\end{equation}
where $\mathbf{x}$ is the position on a 2D plane, $M$ is the mass of the particles in the condensate, $V$ is an external trapping potential and $g$ is the interaction parameter dependent only on 2-point collisions between particles \cite{fetter2001vortices}.
The GPE can be derived from the following action,
\begin{equation} \label{S_GPE}
\mathcal{S}_\mathrm{GPE} = \int dt d^2\mathbf{x} \left[ \frac{i\hbar}{2}\left(\dot{\Psi}\Psi^*-\dot{\Psi}^*\Psi\right)-\frac{\hbar^2}{2M}\bm{\nabla}\Psi\cdot\bm{\nabla}\Psi^* - V|\Psi|^2 - \frac{1}{2}g|\Psi|^4\right],
\end{equation}
where overdot denotes the derivative with respect to time and $\Psi^*$ is the complex conjugate of the wavefunction.

Under the Madelung transform,
\begin{equation}
\Psi(t,\mathbf{x}) = \sqrt{\rho(t,\mathbf{x})}e^{-i\Theta(t,\mathbf{x})/\hbar},
\end{equation}
the GPE reduces to the equations for an irrotational fluid flow,
\begin{subequations} 
\begin{align}
\partial_t\rho + \bm{\nabla}\cdot(\rho\mathbf{v}) = & \ 0, \label{fluid_a} \\
\tfrac{1}{2}M\mathbf{v}^2 + g\rho + V(\mathbf{x})+V_\mathrm{Q} = & \ \mu, \label{fluid_b}
\end{align}
\end{subequations}
with $\rho$ the fluid density, $\mathbf{v} = \bm{\nabla}\Phi$ the velocity field and $\Phi=-\Theta/M$ the velocity potential.
The fluid is assumed stationary so that $\partial_t\Theta=\mu=\cst$, where $\mu$ is the chemical potential associated with the removal of a particle from the condensate.
Equations \eqref{fluid_a} and \eqref{fluid_b} are almost identical to the classical equations except for the addition of the quantum pressure term,
\begin{equation}
V_\mathrm{Q} = -\frac{\hbar^2}{2M}\frac{\nabla^2\sqrt{\rho}}{\sqrt{\rho}}.
\end{equation}
Note also that \eqref{fluid_a} is simply the conserved current associated with the internal symmetry of \eqref{S_GPE}.
The conserved quantity $N=\int d^2\mathbf{x} ~\rho$ gives the number of particles in the condensate.

\subsection{Velocity field} \label{sec:velocity}

Consider now the form of the velocity field.
I will assume that $\mathbf{v}$ is independent $\theta$ and that the density is approximately uniform, i.e $\rho\approx\cst$.
In this case, \eqref{fluid_a} becomes $\bm{\nabla}\cdot\mathbf{v}=0$ and \eqref{fluid_b} implies $\bm{\nabla}\times\mathbf{v}=0$.
The unique velocity profile for a $\theta$ independent fluid is then,
\begin{equation} \label{DBT}
\mathbf{v} = -\frac{D}{r}\vec{\mathbf{e}}_r + \frac{C}{r}\vec{\mathbf{e}}_\theta,
\end{equation}
where $C$ and $D$ are constants.
In this work, I will be interested in draining profiles, hence $D$ is taken to be positive.
$C$ is the circulation parameter which can be either positive or negative depending on the direction of rotation (here I choose $C>0$).
Since $\Psi$ must be periodic in $\theta$ to satisfy the boundary conditions, $C$ must be of the form,
\begin{equation} \label{circ}
C = \hbar \ell/M,
\end{equation}
where $\ell$ is an integer called the winding number.
This is the well-known result that circulation in a BEC is quantised \cite{pitaevskii1961vortex}.
The flow profile in \eqref{DBT} is known as the DBT in the literature.
Note that vortices with winding number higher than $\ell=1$ are usually unstable \cite{shin2004dynamical} and in fact, this instability has been argued to be related to the presence of an ergoregion (the same mechanism responsible for superradiance) \cite{giacomelli2020ergoregion}.
Stabilisation mechanisms, e.g. via trapping potentials, have however been demonstrated \cite{lundh2002multiply,huhtamaki2006dynamically}.

In classical fluids, the angular component $v_\theta=\vec{\mathbf{e}}_\theta\cdot\mathbf{v}$ in \eqref{DBT} is often used as an idealisation of realistic velocity profiles \cite{torres2020quasinormal}.
In the present case, however, it is the true (and only) form of the angular velocity profile for an axisymmetric system.
For the radial profile $v_r=\vec{\mathbf{e}}_r\cdot\mathbf{v}$, one can imagine pumping atoms out of the system near $r=0$ at a rate $\dot{N} = -\rho D$ (the possibility of experimentally realising such a configuration has been discussed in e.g. \cite{bloch1999atom,zezyulin2014stationary}).
In order to keep $N$ fixed, one could then resupply atoms at the same rate at the outer edge of the condensate.

\subsection{Fluctuations}

Now consider fluctuations of the condensate density and phase,
\begin{equation}
\rho\to \rho(1+\eta), \quad \Phi\to \Phi+\phi.
\end{equation}
Linearising \eqref{fluid_a} and~\eqref{fluid_b} in the constant density approximation yields,
\begin{equation} \label{wave_eqn}
\begin{split}
D_t\phi + c^2\eta - \Lambda\nabla^2\eta = & \ 0, \\
D_t\eta + \nabla^2\phi = & \ 0,
\end{split}
\end{equation}
where $D_t=\partial_t+\mathbf{v}\cdot\bm{\nabla}$ is the material derivative and the constants $c$ and $\Lambda$ are given by,
\begin{equation}
c = \sqrt{g\rho/M}, \quad \Lambda = \hbar^2/4M^2.
\end{equation}
For $\Lambda=0$, the system is non-dispersive and all wavelengths will propagate at the same speed $c$.
When $\Lambda\neq 0$, shorter-wavelengths travel faster than $c$ and the system becomes ``superluminally'' dispersive \footnote{The name derives from the analogy with relativity where $c$ represents the speed of light.}.
Since the system is invariant under a rescaling by two parameters, I will set $c=D=1$ from here on.
The background is then completely characterised by choosing $C$ and $\Lambda$.

Note that the equations in \eqref{wave_eqn} can be derived by minimising the action,
\begin{equation} \label{action}
\mathcal{S} = \int dt d^2\mathbf{x} \left[\tfrac{1}{2}\phi D_t\eta-\tfrac{1}{2}\eta D_t\phi - \tfrac{1}{2}\eta^2 - \tfrac{1}{2}(\bm{\nabla}\phi)^2 - \tfrac{1}{2}\Lambda(\bm{\nabla}\eta)^2\right],
\end{equation}
where the term in square brackets is the Lagrangian density $\mathcal{L}$.

\subsection{Conserved currents}

By Noether's theorem, symmetries of the action give rise to conserved currents \cite{schwartz2014quantum}.
In particular, the transformation,
\begin{equation}
\phi \to \phi + \delta \phi,
\end{equation}
and similarly for $\eta$, is called a symmetry if the corresponding change in the Lagrangian can be written in the form $\delta\mathcal{L} = \partial_t f + \bm{\nabla}\cdot\mathbf{F}$, since this leaves $\mathcal{S}$ invariant.
When the equations of motion are satisfied, $\delta\mathcal{L}$ is given by,
\begin{equation}
\delta\mathcal{L} = \partial_t\left(\frac{\partial\mathcal{L}}{\partial\dot{\phi}}\delta\phi + \frac{\partial\mathcal{L}}{\partial\dot{\eta}}\delta\eta\right)+ \bm{\nabla}\cdot\left(\frac{\partial\mathcal{L}}{\partial\bm{\nabla}\phi}\delta\phi+\frac{\partial\mathcal{L}}{\partial\bm{\nabla}\eta}\delta\eta \right).
\end{equation}
Combining these two forms for $\delta\mathcal{L}$, gives the following conservation law,
\begin{equation}
\partial_t\rho + \bm{\nabla}\cdot \mathbf{J} = 0,
\end{equation} 
where the components of the current are given by,
\begin{equation}
\begin{split}
\rho[\phi] = & \ \frac{\partial\mathcal{L}}{\partial\dot{\phi}}\delta\phi + \frac{\partial\mathcal{L}}{\partial\dot{\eta}}\delta\eta - f, \\
\mathbf{J}[\phi] = & \ \frac{\partial\mathcal{L}}{\partial\bm{\nabla}\phi}\delta\phi + \frac{\partial\mathcal{L}}{\partial\bm{\nabla}\eta}\delta\eta - \mathbf{F}.
\end{split}
\end{equation}
This $\rho$ (which is the time component of the current) is not to be confused with the density defined earlier.

Due to the $t$ and $\theta$ independence of $\mathbf{v}$, the action will be invariant under $t$ and $\theta$ translations.
The corresponding conservation laws are the conservation of energy and angular momentum respectively.
In what follows, I will be particularly interested in the radial components of these currents.
These are,
\begin{equation} \label{e_curr}
J_E^r[\phi] = \left(-\tfrac{1}{2}v_r\eta-\partial_r\phi\right)\partial_t\phi + \left(\tfrac{1}{2}v_r\phi-\Lambda\partial_r\eta\right)\partial_t\eta,
\end{equation}
for the energy current and,
\begin{equation} \label{l_curr}
J_L^r[\phi] = \left(-\tfrac{1}{2}v_r\eta-\partial_r\phi\right)\partial_\theta\phi + \left(\tfrac{1}{2}v_r\phi-\Lambda\partial_r\eta\right)\partial_\theta\eta,
\end{equation}
for the angular momentum current.

\subsection{Mode decomposition}

Due to the symmetry of background, it is beneficial decompose the fields $\phi$ and $\eta$ into their different frequency $\omega$ and azimuthal $m$ components.
In this paper, I will work with the following notation,
\begin{equation} \label{comp_pos_neg}
\phi = \sum_\lambda \left(\alpha_\lambda\varphi_\lambda + \alpha^*_\lambda\varphi^*_\lambda\right), \qquad \eta = \sum_\lambda \left(\alpha_\lambda n_\lambda + \alpha^*_\lambda n^*_\lambda\right),
\end{equation}
with $\lambda$ denoting a particular $\omega,m,j$ triplet.
The field modes are,
\begin{equation} \label{decomp}
\varphi_\lambda\equiv\varphi_{\omega mj}(t,\theta,r) = \tilde{\varphi}_j(\omega,m,r)e^{im\theta-i\omega t},
\end{equation}
and similarly for $n_\lambda$.
The $\mathbb{C}$-fields $\varphi$ and $\varphi^*$ are often called positive and negative frequency components respectively. 
The $\alpha_\lambda$ are constant amplitudes multiplying the $\mathbb{C}$-fields, which need to be taken in a symmetric combination due to the fact that $\phi$ and $\eta$ are both real.
The sum over $\lambda$ is short for,
\begin{equation}
\sum_\lambda = \sum_{m,j}\int d\omega,
\end{equation}
where the integral runs from $\omega\in[0,\infty)$ and the azimuthal sum is over $m\in(-\infty,\infty)$.
Finally, $\tilde{\varphi}_j$ is a particular solution to the radial equations of motion, which are obtained by substituting \eqref{decomp} into \eqref{wave_eqn},
\begin{equation} \label{rad_eq}
\begin{split}
-i\left(\omega-\frac{mC}{r^2}\right)\tilde{\varphi}_j + \frac{1}{r}\partial_r\tilde{\varphi}_j  + \left(1+\frac{\Lambda m^2}{r^2}\right)\tilde{n}_j - \frac{\Lambda}{r}\partial_r \tilde{n}_j - \Lambda\partial_r^2\tilde{n}_j = & \ 0, \\
-i\left(\omega-\frac{mC}{r^2}\right)\tilde{n}_j + \frac{1}{r}\partial_r\tilde{n}_j -\frac{m^2}{r^2}\tilde{\varphi}_j + \frac{1}{r}\partial_r \tilde{\varphi}_j + \partial_r^2\tilde{\varphi}_j = & \ 0.
\end{split}
\end{equation}
When $\Lambda=0$, these combine into a single second order ordinary differential equation and one will have $j=1,2$.
Conversely, for $\Lambda\neq 0$, \eqref{rad_eq} has four independent solutions, i.e. $j=1,2,3,4$.

Due to the linearity of the equations of motion \eqref{wave_eqn}, each $\lambda$ component evolves independently and can therefore be considered separately.
Similarly, the positive and negative frequency parts will also evolve independently.
The Lagrangian governing the individual field modes is,
\begin{equation}
\begin{split}
\mathcal{L}_\mathbb{C} = \tfrac{1}{2}\big(\tfrac{1}{2}\varphi_\lambda^* D_t n_\lambda & \ + \tfrac{1}{2}\varphi_\lambda D_t n^*_\lambda -\tfrac{1}{2}n^*_\lambda D_t\varphi_\lambda - \tfrac{1}{2}n_\lambda D_t \varphi^*_\lambda \\
& \ -  n_\lambda n^*_\lambda - \bm{\nabla}\varphi_\lambda\cdot\bm{\nabla}\varphi^*_\lambda - \Lambda\bm{\nabla} n_\lambda\cdot\bm{\nabla} n^*_\lambda \big).
\end{split}
\end{equation}
Applying Noether's theorem for the internal symmetry $\varphi_\lambda\to \varphi_\lambda e^{-i\alpha}$ (and also for $\varphi^*_\lambda,n_\lambda,n^*_\lambda$) one finds the conservation of the norm current, whose components are,
\begin{equation} \label{norm_current}
\begin{split}
\rho_N[\varphi] = & \ \frac{i}{2}\left(\varphi_\lambda n^*_\lambda - n_\lambda \varphi^*_\lambda\right), \\
\mathbf{J}_N[\varphi] = & \ \frac{i}{2}\Big\{ \mathbf{v}\left[\varphi_\lambda n^*_\lambda - n_\lambda \varphi^*_\lambda\right] + \varphi_\lambda \bm{\nabla}\varphi^*_\lambda - (\bm{\nabla}\varphi_\lambda)\varphi^*_\lambda \\
& \qquad + \Lambda\left[ n_\lambda \bm{\nabla}n^*_\lambda - (\bm{\nabla}n_\lambda)n^*_\lambda \right] \Big\}.
\end{split}
\end{equation}
This motivates the definition of the following inner product of two functions (which solve the equations of motion),
\begin{equation} \label{norm1}
(\varphi_{\lambda_1},\varphi_{\lambda_2}) = \frac{i}{2}\int d^2\mathbf{x} \left(\varphi_{\lambda_1} n^*_{\lambda_2} - n_{\lambda_1} \varphi^*_{\lambda_2}\right).
\end{equation}
Since this is independent of $t$, the following quantity is conserved radially,
\begin{equation} \label{Wronsk}
\begin{split}
W[\varphi_{\lambda_1},\varphi_{\lambda_2}] = & \ \frac{i}{2}r\Big\{ v_r\left[\varphi_{\lambda_1} n^*_{\lambda_2} - n_{\lambda_1} \varphi^*_{\lambda_2}\right] + \varphi_{\lambda_1} \partial_r\varphi^*_{\lambda_2} - (\partial_r\varphi_{\lambda_1})\varphi^*_{\lambda_2} \\
& \qquad + \Lambda\left[ n_{\lambda_1} \partial_r n^*_{\lambda_2} - (\partial_r n_{\lambda_1})n^*_{\lambda_2} \right]   \Big\}.
\end{split}
\end{equation}

\section{WKB solutions} \label{sec:wkb}

In this section, I drop the subscript $\lambda$ to avoid complicating the notation.
It will be restored in later sections where it is necessary.
For the velocity profile in \eqref{DBT}, the equations of motion \eqref{wave_eqn} do not admit closed form solutions.
However, if the fluctuations vary on a scale $k^{-1}$ which is much smaller that the scale $L$ over which $\mathbf{v}$ changes, one can define a small parameter $\epsilon=1/kL \ll 1$ and write,
\begin{equation} \label{WKBansatz}
\varphi = \mathcal{A}(\mathbf{x},t)\exp\left(\frac{iS(\mathbf{x},t)}{\epsilon}\right), \quad n = \mathcal{B}(\mathbf{x},t)\exp\left(\frac{iS(\mathbf{x},t)}{\epsilon}\right),
\end{equation}
where $\mathcal{A}$ and $\mathcal{B}$ are local amplitudes and $S$ is the phase.
The solution is obtained by substituting \eqref{WKBansatz} in \eqref{wave_eqn} and solving order by order in $\epsilon$.
In practice, the first two orders are usually all that is needed to obtain a good approximation.
The approximation improves as the wavelength decreases and $\epsilon$ becomes smaller.
In what follows, this will be the case for large $m$.

\subsection{Dispersion relation}

At $\mathcal{O}(\epsilon^0)$, the equations of motion \eqref{wave_eqn} give the Hamilton-Jacobi equation,
\begin{equation} \label{HamJac1}
\left(\partial_t S + \mathbf{v}\cdot\bm{\nabla}S\right)^2 - (\bm{\nabla} S)^2 - \Lambda(\bm{\nabla} S)^4 = 0.
\end{equation}
Identifying the frequency and wavevector through,
\begin{equation} \label{Def1}
\omega = -\partial_t S, \qquad \mathbf{k} = \bm{\nabla}S,
\end{equation}
with $k=||\mathbf{k}||$, the Hamilton-Jacobi equation is equivalent to the dispersion relation,
\begin{equation} \label{disp}
\Omega^2\equiv \left(\omega-\mathbf{v}\cdot\mathbf{k}\right)^2 = k^2+\Lambda k^4,
\end{equation}
which determines the relationship between the local values of $\omega$ and $\mathbf{k}$ when $\mathbf{v}$ is varying.
Note that for the case of $\mathbf{v}=0$, \eqref{disp} is the Bogoliubov dispersion relation originally derived in \cite{bogoliubov1947super}.
Using this notation, \eqref{wave_eqn} can be used to write a leading order relation between the amplitudes,
\begin{equation}
\mathcal{B} = i\Omega f^{-1}\mathcal{A}, \quad f = 1+\Lambda k^2.
\end{equation}
Since the dispersion relation is quadratic in $\omega$, it has two branches,
\begin{equation} \label{branches}
\omega^\pm_D = \mathbf{v}\cdot\mathbf{k}\pm\sqrt{k^2+\Lambda k^4},
\end{equation}
with $\omega^+_D$ the upper branch and $\omega^-_D$ the lower branch.
The group velocity defines the direction of travel of a mode and is given by,
\begin{equation} \label{group}
\bm{v}_g = \bm{\nabla}_\mathbf{k}\omega = \mathbf{v}\pm \mathbf{k}\frac{1+2\Lambda k^2}{\sqrt{k^2+\Lambda k^4}}.
\end{equation}
This also determines the direction in which energy is carried.

As \eqref{HamJac1} is a first order partial differential equation, its solution can be obtained by first splitting into a system of ordinary differential equations and solving these for characteristic curves.
These characteristics can be found from an effective Hamiltonian $\mathcal{H}$ which, using \eqref{branches}, can be expressed concisely as,
\begin{equation} \label{Hamiltonian}
\mathcal{H} = -\frac{1}{2}(\omega-\omega_D^+)(\omega-\omega_D^-).
\end{equation}
The characteristics are obtained as the solutions of Hamilton's equations,
\begin{equation} \label{HamiltonsEqs}
\dot{x}^\mu = \frac{\partial\mathcal{H}}{\partial k_\mu}, \qquad \dot{k}_\mu = -\frac{\partial\mathcal{H}}{\partial x^\mu}
\end{equation}
where $x^\mu=(\mathbf{x},t)$, $k_\mu=(\mathbf{k},-\omega)$. In this section, the overdot denotes the derivative with respect to $\tau$ which parametrises the characteristics.
Solving the system of equations \eqref{HamiltonsEqs} gives the coordinates and the conjugate momenta in terms of the parameter $\tau$, i.e. $x^\mu=x^\mu(\tau)$ and $k_\mu=k_\mu(\tau)$.
The phase part of $\varphi$ in \eqref{WKBansatz} can then be reconstructed by integrating \eqref{Def1} along the different trajectories.
In addition to \eqref{HamiltonsEqs}, the solutions are required to satisfy the Hamiltonian constraint,
\begin{equation} \label{onshell}
\mathcal{H}=0,
\end{equation}
which guarantees that they lie on one of the two branches of the dispersion relation \eqref{disp}.

The analysis can be simplified by specifying to the $t$ and $\theta$ independent system introduced in Section~\ref{sec:velocity}.
In polar coordinates, the wave vector has components,
\begin{equation} \label{equation_k}
\mathbf{k} = (p,m/r), \qquad k = \sqrt{p^2+m^2/r^2},
\end{equation}
where $p$ is the radial wave vector.
By Hamilton's equations \eqref{HamiltonsEqs}, $\omega$ and $m$ are fixed for a given mode, hence, the only variables appearing in the effective Hamiltonian are $r$ and $p$.
The equation $\mathcal{H}=0$ can then be solved directly for $p=p(r)$, thereby circumventing the need to introduce a parameter $\tau$ and solve \eqref{HamiltonsEqs} for $r=r(\tau)$ and $p=p(\tau)$.
The highest power of $p$ in $\mathcal{H}$ will determine the number of solutions that exist.
From here on, these solutions will be labelled $p^\s$ and throughout this work, an upper index will be used to indicate that a particular quantity is associated to the $\s$ solution of the dispersion relation \footnote{Except when discussing the branches of the dispersion relation in \eqref{branches} in which case the superscript $\pm$ indicates the upper and lower branches respectively.}.

\subsection{Transport equation}

At $\mathcal{O}(\epsilon^1)$, the equations of motion \eqref{wave_eqn} give a transport equation for the amplitude,
\begin{equation} \label{transport}
\partial_t(f^{-1}\Omega\mathcal{A}^2) + \bm{\nabla}\cdot(\bm{v}_g f^{-1}\Omega\mathcal{A}^2) = 0,
\end{equation}
which can be solved for $\mathcal{A}$ using the solutions of the Hamilton-Jacobi equation \eqref{HamJac1}.
This equation describes how the amplitude evolves adiabatically along the characteristics.
Using the $t$ and $\theta$ symmetric system of Section~\ref{sec:velocity}, the amplitude is simply,
\begin{equation} \label{amplitude}
\mathcal{A} = |qr|^{-\frac{1}{2}}\mathcal{N}, \qquad q\equiv q(r,p^\s) = f^{-1}(r,p^\s)\mathcal{H}'(r,p^\s),
\end{equation}
where $\mathcal{N}$ is a constant. 
I have also used $\vec{\mathbf{e}}_r\cdot\bm{v}_g\Omega = \mathcal{H}'$ where the prime denotes derivative with respect to $p$.
Hence, the general expression for the radial part of the mode becomes,
\begin{equation} \label{general_mode}
\tilde{\varphi}_j = \sum_\s |q^\s r|^{-\frac{1}{2}} \mathcal{N}^\s_j e^{i\int p^\s dr},
\end{equation}
where the sum over $\s$ accounts for the fact that a given solution of the radial equation may be a combination of WKB modes, and the constants $\mathcal{N}^\s_j$ will be different for each of the $j$ independent solutions.

The norm of an individual $\s$ WKB mode is obtained from \eqref{norm_current} as,
\begin{equation}
\rho_N[\tilde{\varphi}_j^\s] = \frac{\Omega^\s}{f^\s}|\tilde{\varphi}_j^\s|^2,
\end{equation}
and the energy density is simply the same quantity multiplied by $\omega$.
Since $f>0$ for propagating waves, the $\omega>0$ modes with negative energy are those which lie on the $\omega_D^-$ of the dispersion relation where $\Omega<0$.

Next, inserting the full expression \eqref{general_mode} into \eqref{Wronsk} gives, 
\begin{equation}
\begin{split}
W[\tilde{\varphi}_j,\tilde{\varphi}_{j'}] = & \ \frac{1}{2}\sum_{\s,\s'}\Bigg[v_r\left(\frac{\Omega^\s}{f^\s}+\frac{\Omega^{\s'}}{f^{\s'}}\right) + (p^\s+p^{\s'})\left(1+\Lambda\frac{\Omega^\s\Omega^{\s'}}{f^\s f^{\s'}}\right)\Bigg]\frac{\mathcal{N}^\s_j{\mathcal{N}^{\s '}_{j'}}^*}{|q^\s q^{\s '}|^\frac{1}{2}} e^{i\int(p^\s-p^{\s'})dr},
\end{split}
\end{equation}
where the sum is performed over all pairings of modes contained in the different solution.
In this notation, $\tilde{\varphi}_{j'}$ can be a different independent solution to the radial equation for the same $m,\omega$, as encoded by the different set of coefficients $\mathcal{N}^{\s '}_{j'}$.

Since \eqref{Wronsk} is constant in $r$ by definition, and the phase term will cause oscillations if $\s\neq \s'$, the factor in square brackets must vanish for these cases.
This is proven in \cite{coutant2016imprint} for the case of weakly dispersive gravity waves, which also obey a quartic dispersion relation.
Using \eqref{disp}, the factor in square brackets simplifies for $\s=\s'$ and one finds,
\begin{equation} \label{Wronsk2}
W[\tilde{\varphi}_j,\tilde{\varphi}_{j'}] = \sum_l \mathrm{sgn}(q^\s)\mathcal{N}^\s_j{\mathcal{N}^\s_{j'}}^* = \cst.
\end{equation}
This is the key relation from which one can deduce the existence of superradiance in the system, and is equivalent to the energy current up to a factor of $\omega$.


\subsection{Mode mixing}

It is important to note that if the WKB solutions are everywhere valid, then each mode will evolve adiabatically along $r$ without exchanging energy with any of the others (in this case, \eqref{Wronsk2} is trivially satisfied owing to constancy of the $\mathcal{N}_j^\s$).
The locations where the WKB solutions break down thus play an important role in determining the amount of energy exchanged between modes (or mode mixing).
The key assumption underlying WKB is a slowly varying amplitude compared to the phase, hence, the worst possible violation of the approximation occurs when the amplitude suddenly diverges.
Using \eqref{amplitude}, one can see that this occurs if $\mathcal{H}'=0$ somewhere in the system.
From Hamilton's equations \eqref{HamiltonsEqs}, this is equivalent to $\dot{r}=0$.
In other words, these are the locations where an analogous classical particle with energy-momentum relation \eqref{disp} comes to a halt and reverses it's direction, i.e. the classical turning points.
Denoting these locations $r_\tp$, they are found by solving the simultaneous equations,
\begin{equation} \label{TPs}
\mathcal{H}_\tp=0, \qquad \partial_p\mathcal{H}_\tp=0,
\end{equation}
where the subscript $\tp$ denotes that a quantity has been evaluated on a turning point.
Solving these equations yields the pair $(r_\tp,p_\tp)$, i.e. the location of the turning point and the local momentum there.

The turning points also have a simple interpretation in terms of the dispersion relation.
Using \eqref{Hamiltonian}, the conditions in \eqref{TPs} are equivalent to,
\begin{equation}
\omega = \omega_D^\pm(r_\tp,p_\tp), \qquad \partial_p\omega_D^\pm(r_\tp,p_\tp)=0,
\end{equation}
and thus, the turning points are the extrema of the dispersion relation in the $p$ direction.
It is then easy to see why the $r_\tp$ are related to mode mixing.
Consider two $p^\s$ which are initially distinct solutions of the dispersion relation.
As $r$ is varied (and the $\omega_D^\pm$ change shape) the two solutions can approach one another if there is an extremum in between them.
When both solutions sit on the extremum, they have equal $p$ and moving past the turning point, the two modes move off in the complex plane.
In other words, a turning point converts two real solutions of \eqref{onshell} into complex solutions, and in doing so facilitates an interaction between them.

To overcome the breakdown of WKB at turning points, there is an established technique in the literature based on a matched asymptotic expansion.
This method is described fully in e.g. \cite{patrick2020superradiance,patrick2020quasinormal}.
The spirit of the calculation is to expand $\mathcal{H}$ around the turning point, promote this to a wave equation and then write down an exact solution (which turns out to be a combination of Airy functions).
Next, one looks at the asymptotic form of the solution far away from the turning point and notices that this is simply a particular combination of WKB modes.
However, if the asymptotic solution is approached rapidly then one can simply compare the WKB amplitudes at the turning point itself.
This method improves as $m$ increases since the argument of the Airy function grows with $m$, which means it's asymptotic value becomes a better approximation closer to the turning point.
The matrix which relates the WKB modes either side of $r_\tp$ is,
\begin{equation} \label{trans1}
\begin{pmatrix}
A^\OU \\ A^\IN
\end{pmatrix} = T \begin{pmatrix}
A^\downarrow \\ A^\uparrow
\end{pmatrix}, \qquad T = e^{\frac{i\pi}{4}}\begin{pmatrix}
1 & -\frac{i}{2} \\ -i & \frac{1}{2}
\end{pmatrix},
\end{equation}
when the modes are real for $r<r_\tp$ and complex for $r>r_\tp$ and,
\begin{equation} \label{trans2}
\begin{pmatrix}
A^\uparrow \\ A^\downarrow
\end{pmatrix} = \widetilde{T} \begin{pmatrix}
A^\OU \\ A^\IN
\end{pmatrix}, \qquad \widetilde{T} = e^{\frac{i\pi}{4}}\begin{pmatrix}
\frac{1}{2} & -\frac{i}{2} \\ -i & 1
\end{pmatrix},
\end{equation}
when the modes are complex for $r<r_\tp$ and real for $r>r_\tp$.
Here, the propagating modes $\OU$ and $\IN$ are defined so that $p^\OU>p^\IN$.
The complex modes are defined so that $\uparrow$ is the one which grows in the direction of increasing $r$ and $\downarrow$ decays.

\subsection{Transfer matrix}

To relate the WKB amplitudes in the asymptotic regions of the flow, one can define an $M\times M$ matrix (where $M$ is the number of modes in the system) called the transfer matrix, $\mathcal{M}$.
Before writing down $\mathcal{M}$, it will be instructive to establish some preliminaries.

Firstly, since I will ultimately be interested in relations between the different mode amplitudes as determined by \eqref{Wronsk} (which includes a factor of $r$ out the front) it is useful to define a new set of WKB modes,
\begin{equation}
R(r) = \sum_\s A^\s(r) e^{i\int p^\s(r) dr},
\end{equation}
which are related to those in \eqref{general_mode} through $R(r)=\sqrt{r}\tilde{\varphi}(r)$ (note that I have dropped the subscript $j$, denoting the solution to the radial equation, since subscripts in this section will be used to indicate the $r$ location where a quantity is evaluated).
The $\sqrt{r}$ factors out the part of the amplitude which increases simply due to the fact that a wave moving in the direction of decreasing $r$ gets focussed onto a smaller disk.

Now, define a column vector $\mathbf{A}$, which consists of the WKB amplitudes $A^\s$, and a row vector $\mathbf{P}$, containing the WKB phases $e^{i\int p^\s dr}$.
Let's say that we know the full details of the amplitudes and phases at a point $r_b$ and we want to transport this solution to another point $r_a<r_b$ where the WKB approximation holds everywhere along the path.
First, the full solution at $r_b$ is given by $R_b=\mathbf{P}_b\cdot\mathbf{A}_b$.
Then, defining the factor,
\begin{equation}
\mathcal{F}^\s_{ab} = \left|\frac{q^\s_b}{q^\s_a}\right|^\frac{1}{2}\exp\left(-i\int^{r_b}_{r_a}p^\s dr\right),
\end{equation}
the amplitudes can be transported as,
\begin{equation} \label{transp1}
\mathbf{A}_a = \mathrm{diag}(\mathcal{F}_{ab}^\s) \mathbf{A}_b,
\end{equation}
so that the solution at $r_a$ can be defined with respect to original phase vector through $R_a=\mathbf{P}_b\cdot\mathbf{A}_a$.

The transfer matrix $\mathcal{M}$ relates the mode amplitudes in the asymptotic regions,
\begin{equation}
\mathbf{A}_0 = \mathcal{M} \mathbf{A}_\infty,
\end{equation}
where in defining these regions, it suffices (at the considered level of approximation) to find two locations $r_{0,\infty}$ such that there are no turning points for $r<r_0$ or $r>r_\infty$, since the energy content of the different modes beyond these points is then fixed.
With this definition, one can use the solution at $r_\infty$, i.e. $R_\infty=\mathbf{P}_\infty\cdot\mathbf{A}_\infty$, to deduce the same at $r_0$, i.e. $R_0=\mathbf{P}_\infty\cdot\mathbf{A}_0$ with $\mathbf{A}_0$ given above.
To construct $\mathcal{M}$, one performs a series of matrix multiplications using \eqref{trans1}, \eqref{trans2} and \eqref{transp1} (see \cite{patrick2020quasinormal} for an explicit example of this).
However, due to the way that $T$ and $\widetilde{T}$ act on the amplitude vectors, the situation is a bit different if there are two turning points (say $r_a<r_b$) where two real modes are converted into complex modes and then back into real modes.
In this case, the amplitudes of the interacting modes are related via,
\begin{equation} \label{LocalScatter}
\begin{pmatrix}
A_a^\OU \\ A_a^\IN
\end{pmatrix} = \mathcal{N}_{ab} \begin{pmatrix}
A_b^\OU \\ A_b^\IN
\end{pmatrix},
\end{equation}
with,
\begin{equation}
\begin{split}
\mathcal{N}_{ab} = & \ \mathcal{F}_{ab}^\downarrow\begin{bmatrix}
1+\tfrac{1}{4}f_{ab}^2 & i\left(1-\tfrac{1}{4}f_{ab}^2\right) \\ -i\left(1-\tfrac{1}{4}f_{ab}^2\right) & 1+\tfrac{1}{4}f_{ab}^2
\end{bmatrix}, \\
f_{ab} = & \ \exp\left(-\int^{r_b}_{r_a}\mathrm{Im}[p^\downarrow]dr\right),
\end{split}
\end{equation}
where the $\downarrow$ mode is the complex solution of the dispersion relation which decays with increasing $r$ between the two turning points (see \cite{patrick2020quasinormal} for details).

For the scattering problems considered in this work, it turns out that \eqref{LocalScatter} contains all the necessary physics to compute the amount of superradiance.
This is because, in the two mode case (i.e $\Lambda=0$) one can define $r_0=r_a$ and $r_\infty=r_b$ and then \eqref{LocalScatter} is equivalent to the full transfer matrix.
In the four mode case (i.e $\Lambda\neq 0$) the modes decouple into two pairs (i.e. $\mathcal{M}$ becomes block diagonal) and the computation of the scattering coefficients proceeds identically to the two mode case.

\section{Non-dispersive modes in the DBT} \label{sec:nondisp}

\begin{figure} 
\centering
\includegraphics[width=.6\linewidth]{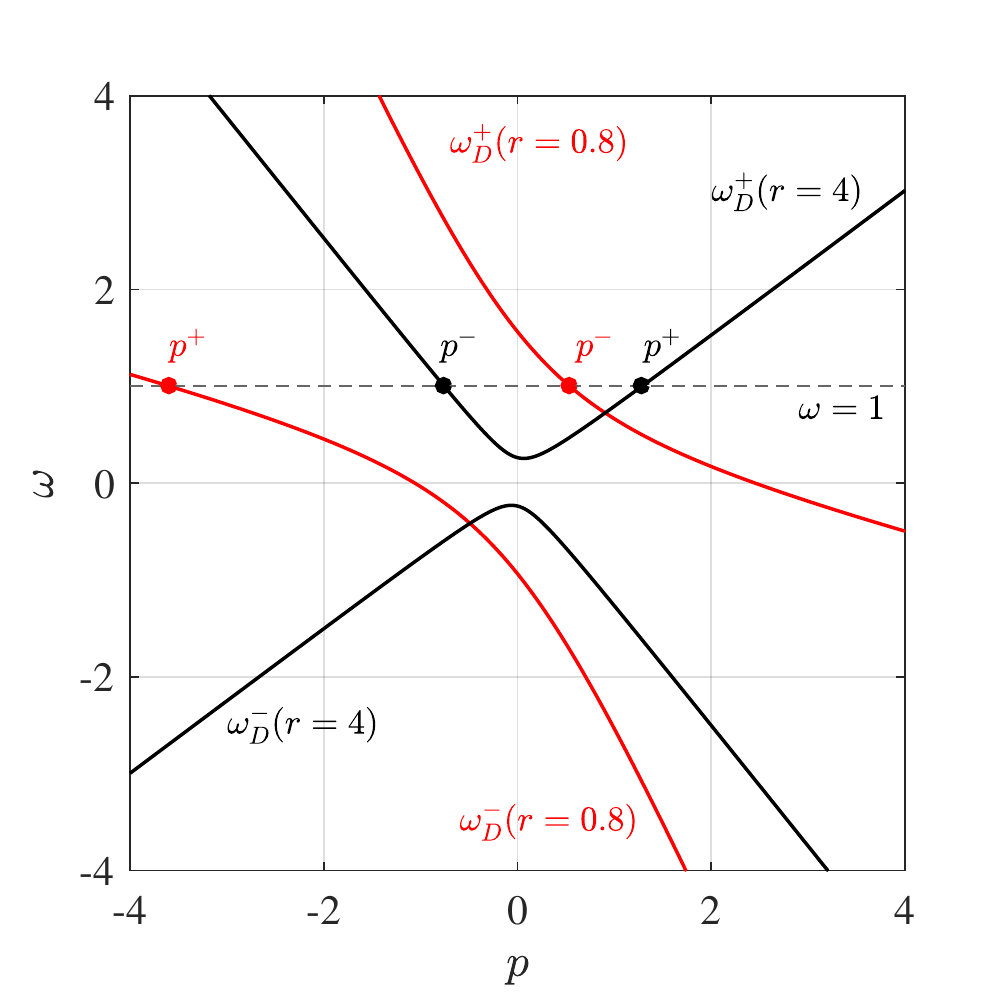}
\caption{An example of the branches of the dispersion relation \eqref{branches} for $m=1$ and $C=0.2$.
Two $r$ locations are shown; one outside the horizon (black curves) and one inside (red curves).
All modes inside the horizon are in-going since the gradient of both branches (in $p$) is everywhere negative.
This particular set of parameters corresponds to Type $\mathrm{II}^0$ scattering, defined further down.
} \label{fig:disprel_nondisp}
\end{figure}

In the case where $\Lambda=0$, the equations of motion \eqref{wave_eqn} can be recast as a Klein-Gordon equation for the fluctuations of a scalar field propogating through an effective $(2+1)$-dimensional spacetime.
This is the conventional way that superradiance in the DBT is studied and has been discussed on many occasions in the literature, e.g. \cite{basak2003reflection,berti2004qnm,basak2005analog}.
Hence, I will not reproduce the analysis here, opting instead to infer the important properties of the system (in particular, the existence of a horizon and superradiance) directly from the dispersion relation.
This approach has the benefit that it can be easily generalised to the dispersive case, where the description in terms of an effective metric is no longer available.

The effective Hamiltonian for this case is,
\begin{equation} \label{Hamiltonian2}
\mathcal{H} = \tfrac{1}{2}\left[(1-1/r^2)p^2 - 2\tilde{\omega}p/r - \tilde{\omega}^2 + m^2/r^2 \right],
\end{equation}
where I have introduced the frequency in the rotating frame,
\begin{equation}
\tilde{\omega} = \omega - mC/r^2.
\end{equation}
This has roots which are labelled $l\in\{+,-\}$, and are given by,
\begin{equation} \label{p_shal}
p^\pm = \frac{\tilde{\omega}/r\pm\sqrt{-V}}{1-1/r^2},
\end{equation}
with $V$ defined by,
\begin{equation} \label{eff_potential}
V = -\tilde{\omega}^2 + (1-1/r^2)m^2/r^2.
\end{equation}
The solutions are labelled such that $\mathrm{Re}[p^+] > \mathrm{Re}[p^-]$ outside of the horizon (to be defined shortly).
In particular, the $+$ mode is radially out-going (i.e. $\vec{\mathbf{e}}_r\cdot\bm{v}_g^+>0$) as $r\to\infty$, whereas the $-$ mode is in-going (i.e. $\vec{\mathbf{e}}_r\cdot\bm{v}_g^-<0$).
Note that when $V>0$, the two roots become complex with identical real parts, and equal and opposite imaginary parts.
In this region they are labelled $l\in\{\uparrow,\downarrow\}$, where the $\uparrow$ mode is the one which grows with increasing $r$, $\mathrm{Im}[p^\uparrow]<0$, and the $\downarrow$ mode is the one which decays with increasing $r$, $\mathrm{Im}[p^\downarrow]>0$.
An example of the functions $p^\pm(r)$ can be found in Fig.~2 of \cite{patrick2020superradiance}, which will shortly be represented in a ``Feynman'' diagrammatic form.

\subsection{The horizon}

Consider now the branches of the dispersion relation in \eqref{branches} as a function of $p$ at a given $r$ (see e.g. Fig.~\ref{fig:disprel_nondisp}).
The intersection of a line of constant $\omega$ with the branches gives the two roots, and by \eqref{group}, the gradient $\partial_p\omega_D^\pm$ at these points gives the local group velocity of the modes.
As $r\to\infty$, the $-$ is always in-going with $\partial_p\omega_D^+(p^-)<0$, whereas the $+$ mode is out-going with $\partial_p\omega_D^+(p^+)>0$.
However, approaching the origin, both modes have $\partial_p\omega_D^+(p^\pm)<0$ and are therefore in-going.
The transition between these two scenarios occurs as the two branches of the dispersion relation rotate clockwise in the $(p,\omega)$ plane and one of the modes is sent to $p^\pm\to\pm\infty$.
Looking at the expression for $p$ in \eqref{p_shal}, this occurs where the denominator is zero, which occurs for $r=r_h\equiv 1$ (or in dimensional units $r_h=D/c$).
This location (the horizon) is the boundary of the region inside of which there are no modes which escape to spatial infinity.

\begin{figure} 
\centering
\includegraphics[width=\linewidth]{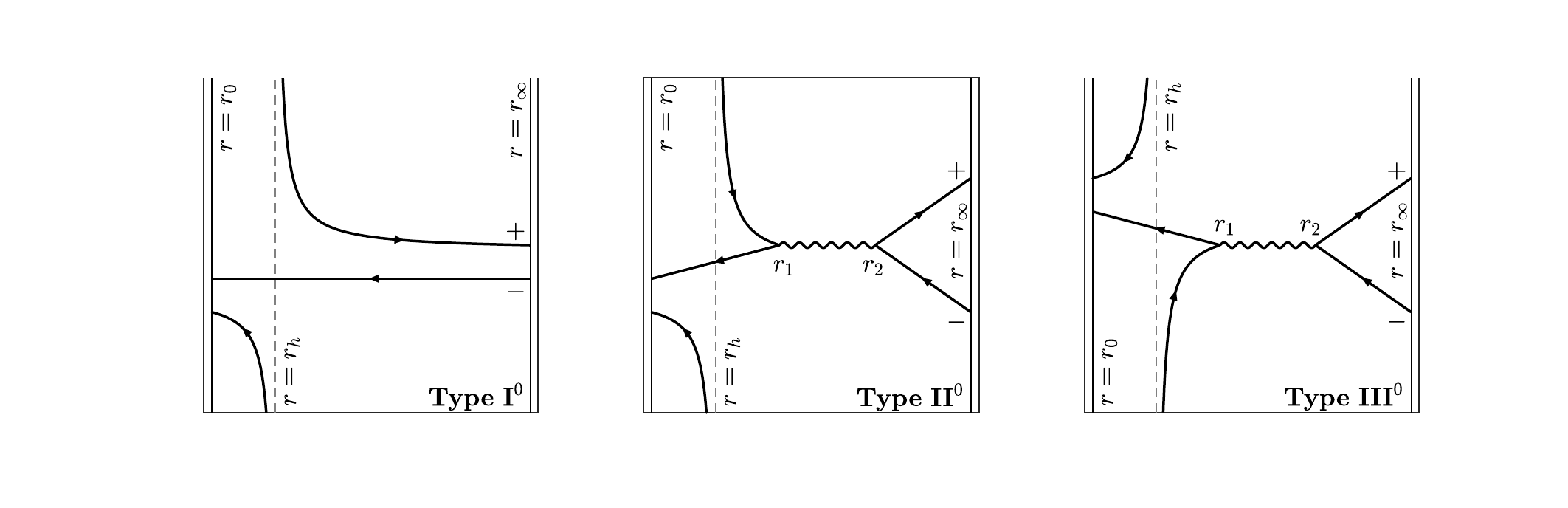}
\caption{
An illustration of the scattering of the two non-disperive modes through the $(r,p)$ phase space in the form of ``Feynman'' diagrams ($r$ on the horizonal and $p$ on the vertical).
Only the real part of $p$ is shown, and increases moving up the diagram.} \label{fig:feynman1}
\end{figure}

\subsection{Scattering types} \label{sec:type_NDSP}

The different scattering possibilities can be classified using a similar scheme to that developed in \cite{patrick2020superradiance}.
In the non-dispersive case, there are three different scattering types which can be represented using phase space diagrams. 
The diagrams involve a schematic illustration of the real part of $p^\pm(r)$ through the $(r,p)$ phase space and take on a similar form to Feynman diagrams.
The important features of these diagrams are the number of turning points, the modes which interact there and the asymptotics of the modes.
For any possible combination of the wave and background parameters, i.e. $\omega,m$ and $C$, the scattering outcome falls into one of the three categories.
Which category it falls into depends on the size of $\omega$ relative to the characteristic frequencies of the system, which are defined now.

The simplest possible outcome is that each WKB mode evolves adiabatically across the system without interacting with the other, i.e. there are no turning points. 
In this case, the $+$ mode diverges at $r_h$ whereas the $-$ mode is regular there.
This will be called Type $\mathrm{I}^0$ scattering\footnote{Note, the superscript $0$ indicates the non-dispersive case, i.e. $\Lambda=0$. In the dispersive case, the superscript $+$ will be used for $\Lambda>0$.}.

The next possibility is that the the modes have the same asymptotics as the previous case, but now there is an interaction between both modes somewhere for $r>r_h$.
Due to the asymptotics, there must be two turning points; one to convert the real modes at large $r$ into complex modes, and a second to convert complex modes back into real modes near the horizon.
In this intermediate region, the complex nature of $p$ leads to an exponential fall off of the amplitude, i.e. the modes \textit{tunnel} between the turning points.
This case will be called Type $\mathrm{II}^0$ scattering.

The transition between Types $\mathrm{I}^0$ and $\mathrm{II}^0$ occurs when the two turning points meet at a single location.
On the dispersion relation, this means that $p^+$ and $p^-$ become equal momentarily before departing back in the direction they came from.
At this point (which is the well known light-ring $r_\lr$ from black hole physics \cite{cardoso2009geodesic}) the condition $\partial_r\omega_D^+=0$ is also satisfied.
Using \eqref{Hamiltonian}, the conditions for this location are,
\begin{equation} \label{LRconds}
\mathcal{H}_\lr = 0, \qquad \partial_r\mathcal{H}_\lr = 0, \qquad \partial_p\mathcal{H}_\lr = 0,
\end{equation}
which yields a triplet $(r_\lr,p_\lr,\omega_\lr)$.
These conditions imply a relation between $r_\lr$ and $p_\lr$,
\begin{equation} \label{pr_cond}
p_\lr r_\lr = B_\pm \equiv mC\pm\sqrt{m^2C^2+m^2},
\end{equation}
where the $+$ sign is for the upper branch and $-$ sign for the lower one.
Note that this relation also holds for $\Lambda\neq 0$.
The light-ring is given by,
\begin{equation}
r^\pm_\lr = \pm(B_\pm^2+m^2)^\frac{1}{2}/B_\pm,
\end{equation}
and the light-ring frequency by,
\begin{equation}
\omega^\pm_\lr = mC/r_\lr^2+(1-1/r_\lr^2)B_\pm.
\end{equation}
Since I consider $\omega>0$ modes, only the light-ring on the upper branch is required and I will therefore set $r_\lr=r^+_\lr$ and $\omega_\lr=\omega^+_\lr$ from here on.
Note that $r_\lr$ is independent of $m$ whereas $\omega_\lr$ scales linearly with $|m|$.

In the final possibility, the turning point structure is the same as the last case but now the $p^+$ mode is regular at $r_h$ with the $p^-$ mode diverging there.
By considering how the dispersion relation evolves with $r$, e.g. in Fig.~\ref{fig:disprel_nondisp}, it is easy to convince oneself that this occurs when the modes are on the lower branch just outside the horizon.
To identify the relevant frequency controlling when this occurs, consider the following.
For a tunnelling mode to re-emerge on either $\omega_D^+$ or $\omega_D^-$, the two branches must have extrema.
However, the extrema vanish at $r_h$ since the right (left) part of the upper (lower) branch asymptotes to,
\begin{equation}
\omega_\star = mC/r_h^2 \equiv m\Omega_h^\mathrm{rot}.
\end{equation} 
Hence, for $\omega>\omega_\star$, the modes will be on the upper branch just outside of $r_h$, whereas for $\omega<\omega_\star$ they will be on the lower branch.
Note that $\omega_\star<\omega_\lr$ for all $m$ and $C$.
This $\omega_\star$ is in fact the well-known threshold frequency for superradiance introduced in \eqref{sr_bh}.
I will show precisely why this condition implies superradiance in the next section. 
For now, it serves as a condition for the final type of scattering, which I call Type $\mathrm{III}^0$.

\begin{figure} 
\centering
\includegraphics[width=.6\linewidth]{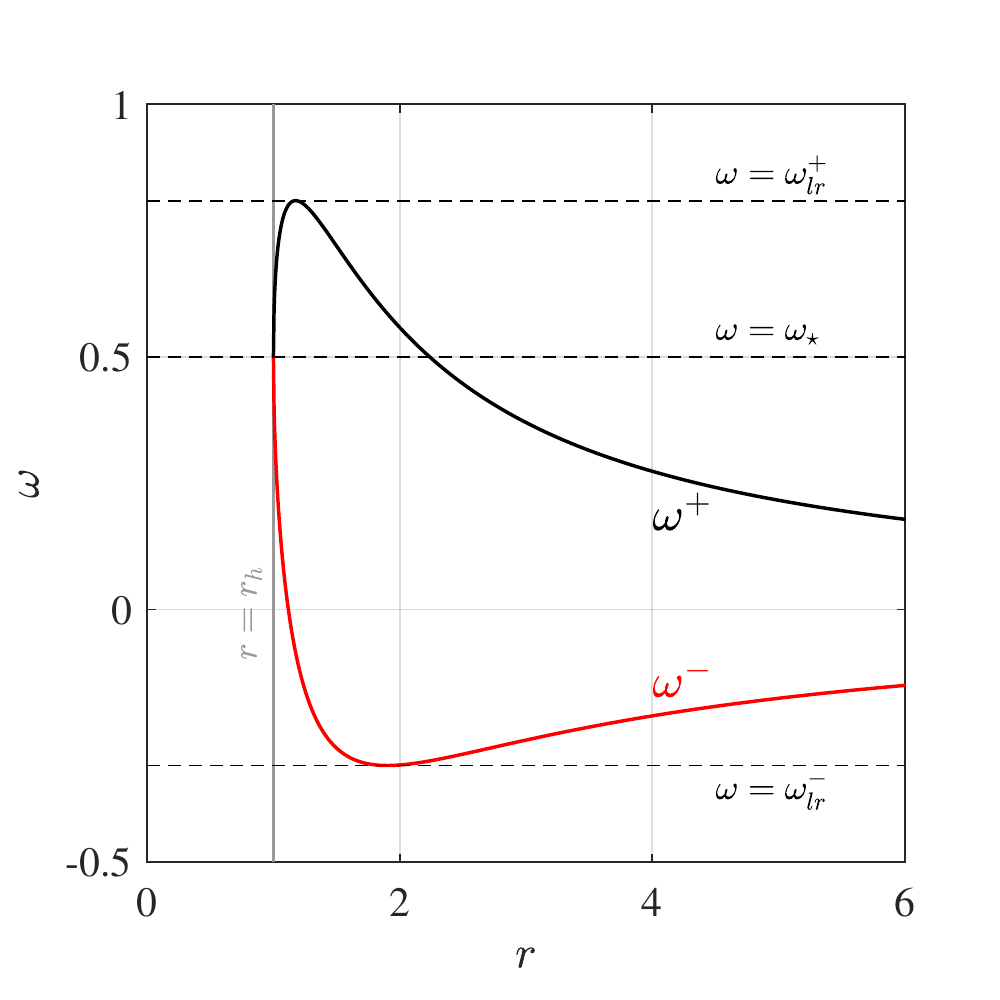}
\caption{The value of $\omega$ at the extrema of $\omega_D^\pm$ for $m=1$, $C=0.5$.
The intersection of a line with $\omega=\cst$ gives the location of the turning points.
Thus, Type $\mathrm{I}^0$ scattering occurs above $\omega^+_{lr}$ and Type $\mathrm{II}^0$ occurs for $\omega_\star<\omega<\omega^+_{lr}$.
Below $\omega_\star$, positive frequencies tunnel to the lower branch of the dispersion relation, which is scattering Type $\mathrm{III}^0$.
Note that since the dispersion relation is invariant under $\{\omega,m\}\to\{-\omega,-m\}$, the scattering of the $m=-1$ mode is described by the same plot inverted about the $r$-axis (the signs on all the labels should also be exchanged).
Thus for $m<0$, Type $\mathrm{II}^0$ occurs for $0<\omega<\omega^+_{lr}$ and Type $\mathrm{I}^0$ for $\omega>\omega^+_{lr}$.
} \label{fig:extrema_nondisp}
\end{figure}

There is another convenient way to understand the different scattering possibilities by plotting the evolution of the extrema of $\omega_D^\pm$ with r.
To do this, one solves $\partial_p\omega_D^\pm=0$ to find a relation $p=p_\ex(r)$, where $+p_\ex$ gives the local momentum at the extrema on the upper branch and $-p_\ex$ gives the same on the lower branch.
Then, the value of $\omega$ on the extrema is given by,
\begin{equation}
\omega^\pm(r)\equiv\omega_D^\pm(r,\pm p_\ex(r))=mC/r^2\pm\sqrt{(1-1/r^2)m^2/r^2}.
\end{equation} 
The turning points can be understood as the intersection of these curves with a line of $\omega=\cst$.
Then $\omega_\lr$ is simply the extremum of $\omega_+$ (for $\omega>0$) in the radial direction and $\omega_\star=\omega^\pm(r_h)$.
These curves are illustrated for a particular value of $m$ and $C$ in Fig.~\ref{fig:extrema_nondisp}.
Note finally that since the function $V$ defined in \eqref{eff_potential} can be written,
\begin{equation}
V=-(\omega-\omega^+)(\omega-\omega^-),
\end{equation}
the turning points correspond to the zeros of $V$.
Hence, $V$ can be thought of as an effective potential barrier.


\subsection{Scattering coefficients} \label{sec:scat_coefs}

Since the radial equation of motion is second order in spatial derivatives, there are two independent solutions $j=1,2$.
Following \cite{unruh1974second}, and using the diagrams in Fig.~\ref{fig:feynman1}, these are defined by the asymptotics,
\begin{equation} \label{asymp}
\begin{split}
R_1 \sim & \  \frac{1}{2\pi|q^-_\infty|^\frac{1}{2}} \times \begin{cases}
e^{i\int^{r_\infty} p^- dr} + \mathcal{R}e^{i\int^{r_\infty} p^+ dr}, \ \ \quad r\to r_\infty, \\ \mathcal{T}e^{i\int^{r_0} p^\mp dr}, \ \ \ \qquad \qquad \qquad r\to r_0,
\end{cases} \\
R_2 \sim & \  \frac{1}{2\pi|q^\pm_0|^\frac{1}{2}} \times \begin{cases}
\mathcal{U}e^{i\int^{r_\infty} p^+ dr}, \qquad \quad \qquad \qquad r\to r_\infty, \\
e^{i\int^{r_0} p^\pm dr} + \mathcal{V}e^{i\int^{r_0} p^\mp dr}, \ \ \qquad  r\to r_0,
\end{cases}
\end{split}
\end{equation}
where the upper sign is taken for Types $\mathrm{I}^0$ and $\mathrm{II}^0$ and the lower sign for Type $\mathrm{III}^0$.
Here, $\mathcal{R},\mathcal{T},\mathcal{U},\mathcal{V}$ are scattering coefficients and the factor of $2\pi$ is there so that the incident part of the mode is normalised in the inner-product \eqref{norm1} (see \ref{app:norm}).
It is also understood that the subscript on $R$ is used to specify the $j$ mode, rather than the radial location as on other quantities.

Plugging these into \eqref{Wronsk2}, the scattering coefficients obey the following relations,
\begin{subequations} \label{scatterfull}
\begin{align}
|q_\infty^-|(2\pi)^2 ~W[\tilde{\varphi}_1,\tilde{\varphi}_1] = & \ q^-_\infty+q^+_\infty|\mathcal{R}|^2 = q^\mp_0|\mathcal{T}|^2, \label{scatter1} \\
|q_0^\pm|(2\pi)^2 ~W[\tilde{\varphi}_2,\tilde{\varphi}_2] = & \ q^+_\infty|\mathcal{U}|^2 = q^\pm_0+q^\mp_0|\mathcal{V}|^2, \\
|q_\infty^-q_0^\pm|^\frac{1}{2}(2\pi)^2~W[\tilde{\varphi}_1,\tilde{\varphi}_2] = & \ q^+_\infty\mathcal{R}\mathcal{U}^* = q^\mp_0\mathcal{T}\mathcal{V}^*,
\end{align}
\end{subequations}
where the factors on the left hand side have been left there for later use when evaluating the quantum currents.

Consider now the classical scattering of an in-coming wave with the vortex (i.e. the $R_1$ solution).
Superradiance occurs when the reflected wave carries away more energy than the incident wave had coming in, i.e. $q^+_\infty|\mathcal{R}|^2>q^-_\infty$, which by \eqref{scatter1} implies that $q_0^\mp<0$ is a necessary (and sufficient) condition for superradiance.
Using $q=f^{-1}\vec{\mathbf{e}}_r\cdot\bm{v}_g\Omega$ and realising that in $R_1$ the solution on the horizon is always in-going, superradiance will occur provided $\Omega_0<0$.
This is true for the solution which tunnels to the lower branch of the dispersion relation, i.e. the one in Type $\mathrm{III}^0$ scattering.

The condition for superradiance assumes a more familiar form if one sets $r_\infty$ to be true spatial infinity and $r_0$ to sit just outside $r_h$.
Then $q_\infty^+=-q_\infty^-=\omega$ and $q_0^\mp=-\tilde{\omega}_h$, and \eqref{scatter1} becomes,
\begin{equation}
|\mathcal{R}|^2+\frac{\tilde{\omega}_h}{\omega}|\mathcal{T}|^2 = 1,
\end{equation}
which is the usual relation between scattering coefficients from black hole physics.
Thus amplification occurs for $\tilde{\omega}_h<0$, which corresponds to Type $\mathrm{III}^0$ scattering.

To find an expression for the reflection coefficient, I will again exploit the freedom to move the points $r_0$ and $r_\infty$ (although identical results are found when these locations are fixed \cite{patrick2020superradiance}).
In Type $\mathrm{I}^0$ scattering, there are no real turning points and thus, at the considered level of approximation, one has $|\mathcal{R}|=0$.
For the other two cases, choose $r_0$ to sit just inside $r_1$, and $r_\infty$ just outside $r_2$.
Applying the formula in \eqref{LocalScatter} and inserting the amplitudes for $R_1$, the reflection coefficient is given by,
\begin{equation} \label{refl}
|\mathcal{R}| = \left(\frac{1-f_{12}^2/4}{1+f_{12}^2/4}\right)^{\mathrm{sgn}(\tilde{\omega}_h)},
\end{equation}
which as expected satisfies $|\mathcal{R}|<1$ for Type $\mathrm{II}^0$ and $|\mathcal{R}|>1$ for Type $\mathrm{III}^0$.
This will be plotted later on in Fig.~\ref{fig:refls} along with the dispersive solutions.

\section{Dispersive modes in the DBT} \label{sec:disp}

\begin{figure} 
\centering
\includegraphics[width=.6\linewidth]{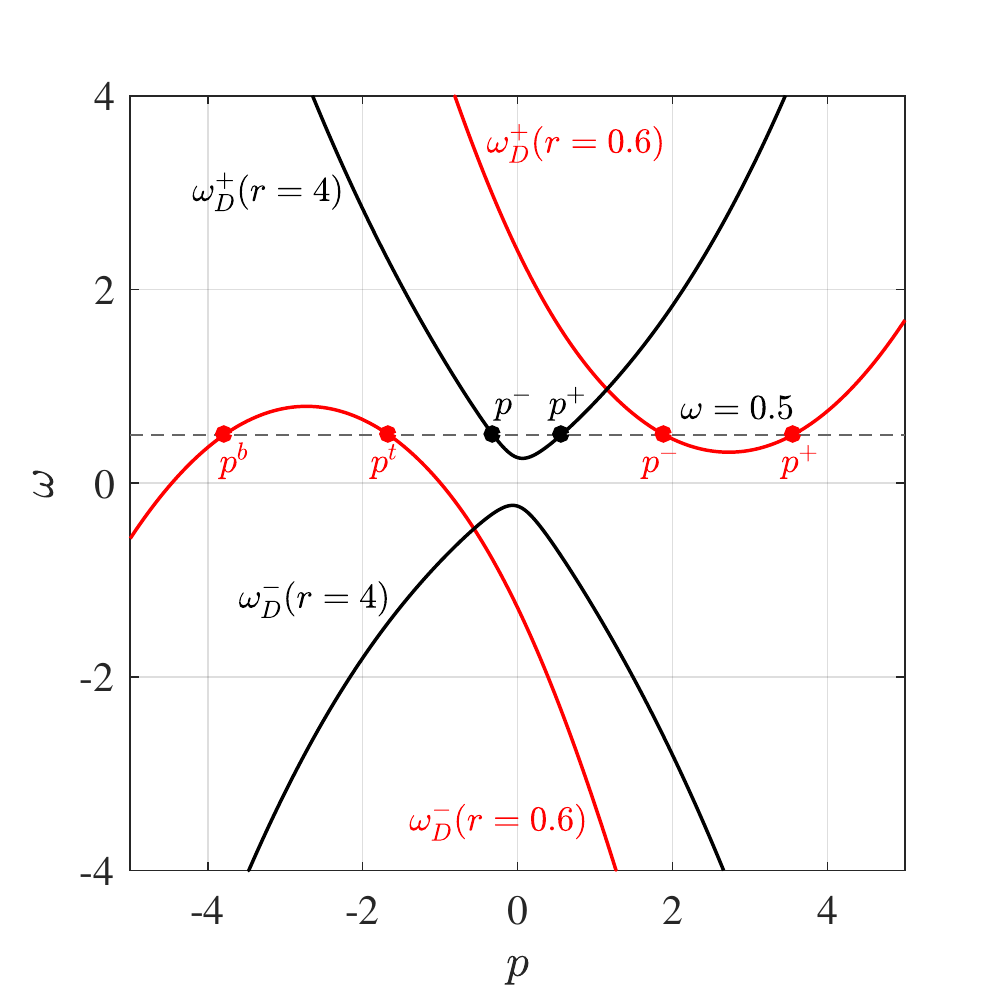}
\caption{An example of the branches of the dispersion relation \eqref{branches} for $m=1$, $C=0.2$ and $\Lambda=0.08$.
At large $r$ (black curves) there are only two real solutions of the dispersion relation, $p^\pm$, which are the same as those in the non-dispersive case.
At small $r$ (red curves) it is possible for the other solutions, $p^{t,b}$, to also become real.
} \label{fig:disprel_disp}
\end{figure}

\begin{figure} 
\centering
\includegraphics[width=.6\linewidth]{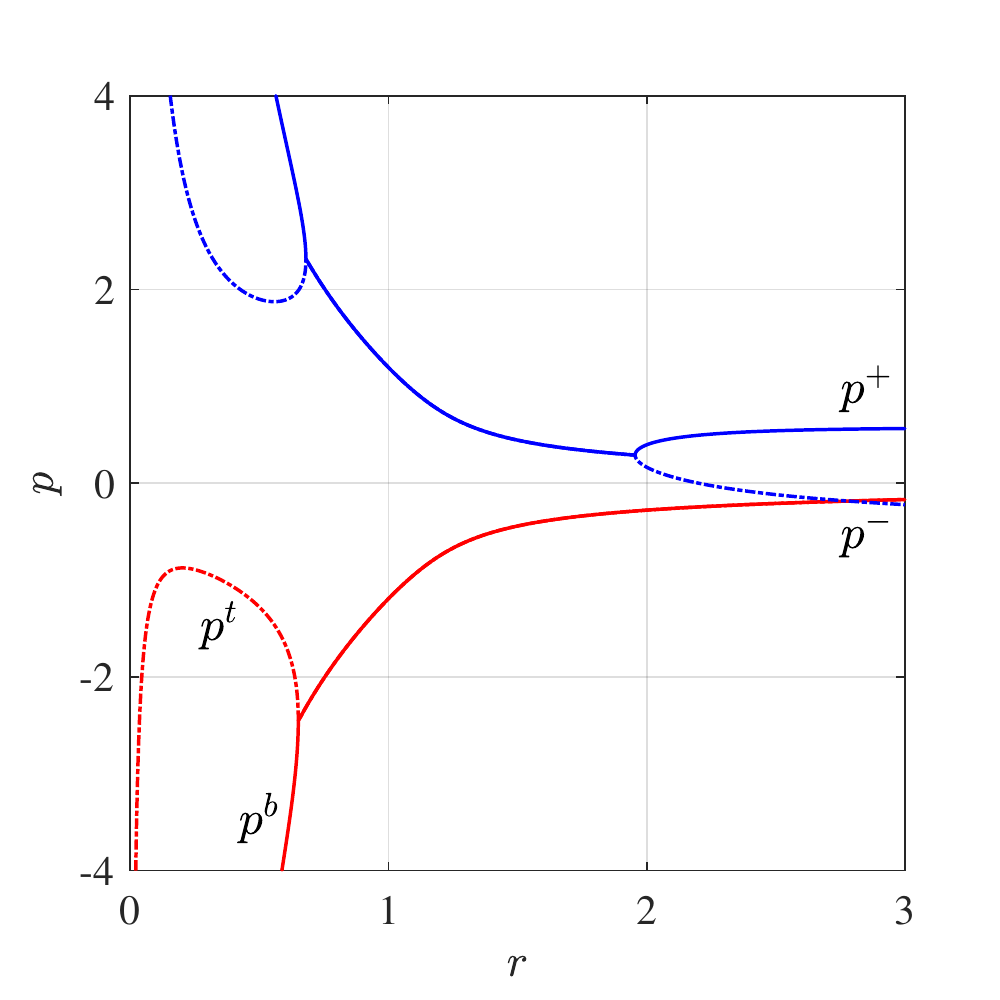}
\caption{The evolution of the four modes with $r$ for the same parameters as in Fig.~\ref{fig:disprel_disp}.
Only the real part of $p$ is shown.
When neighbouring trajectories intersect in phase space, mode mixing occurs.
This particular diagram corresponds to Type $\mathrm{VIII}^+_\mathrm{a}$ scattering defined later on.
Note that the trajectories of the complex $p^{t,b}$ modes at large $r$ do not really intersect that of $p^-$, but rather pass around it in the complex plane.
} \label{fig:mode_trajs}
\end{figure}

The effective Hamiltonian for the dispersive case is given by,
\begin{equation} \label{Hamiltonian4}
\mathcal{H} = \frac{1}{2}\left[\Lambda p^4 + \left(1-\frac{1-2\Lambda m^2}{r^2}\right)p^2 - \frac{2\tilde{\omega}}{r}p -\tilde{\omega}^2 + \frac{m^2}{r^2}\left(1+\frac{\Lambda m^2}{r^2}\right)\right].
\end{equation}
Since this is a depressed quartic, there are now four different solutions which I will label $\s\in\{+,-,t,b\}$.
The $+$ and $-$ solutions are the same ones from the non-dispersive case; in particular, they correspond to the out- and in-going modes as $r\to\infty$ and obey $\mathrm{Re}[p^+]\geq\mathrm{Re}[p^-]$ everywhere.
The $t$ and $b$ modes arise due to dispersion and are defined to be those which satisfy $p^{t,b}(r\to\infty)\in\mathbb{C}$ and $\mathrm{Re}[p^t]\geq\mathrm{Re}[p^b]$.
In many cases, these modes can become real, propagating solutions in the vortex core.
An example of this is given in Fig.~\ref{fig:disprel_disp}, where it is shown how the dispersive modes arise on the dispersion relation.
The functions $p^\s(r)$ for the same parameters are displayed in Fig.~\ref{fig:mode_trajs}.

An important difference between the non-dispersive and dispersive cases is the absence of a horizon in the latter.
The reason for this is that, due the $p^4$ term in \eqref{Hamiltonian4}, there is no longer a critical radius below which $\partial_p\omega_D^\pm<0$ for all $p$ (see Fig.~\ref{fig:disprel_disp} where this behaviour is readily apparent).
Consequently, there is no blocking of out-going high momentum modes approaching the origin. 
(An exception to this is when the propagation of the $+$ and $-$ modes is prohibited in the vortex core. However, when this happens, both in- and out-going modes are blocked rather than just the out-going one). 

\subsection{Scattering types} \label{sec:type_DSP}

The scattering possibilities for $\Lambda>0$ will now be classified in a similar manner to those in Section~\ref{sec:type_NDSP}.
Dispersion significantly enhances the number of possible outcomes compared with the non-dispersive case.
Hence, instead of discussing each case individually, I will instead provide a parameter space plot in Fig.~\ref{fig:paramspace_disp} to illustrate the parameter ranges associated with the different types of scattering.
The parameter space is divided up by three important frequencies which I discuss now.

\subsubsection{Light-ring}

Similarly to the non-dispersive case, the light-ring frequency provides a boundary in parameter space above which the $+$ and $-$ modes decouple.
Using the conditions in \eqref{LRconds} along with $p_\lr r_\lr=B_\pm$ from \eqref{pr_cond}, the location of the light-ring (on $\omega_D^+$) is,
\begin{equation}
r_{lr} = \sqrt{\frac{B_+^2+m^2}{2B_+^2}}\left(1+\sqrt{1-4\Lambda B_+^2}\right)^\frac{1}{2}\left(1-4\Lambda B_+^2\right)^\frac{1}{4}.
\end{equation}
The light-ring momentum and frequency are then immediately given by $p_{lr}=B_+/r_{lr}$ and $\omega_{lr} = \omega_D^+(r_{lr},p_{lr})$.

For the following discussion, it is useful to visualise the light-rings as the extrema of the $\omega^\pm$ curves (which are defined in the same way as in the non-dispersive case, i.e. the value of $\omega$ at the extrema of $\omega_D^\pm$).
I now define the following critical parameters,
\begin{equation} \label{critical_vals}
C_0 = \left|\frac{1-4\Lambda m^2}{4\Lambda^\frac{1}{2}m}\right|, \qquad \Lambda_c = 1/4m^2,
\end{equation}
which play a key role in characterising the scattering.
To give some intuition about the significance of these parameters, I discuss below their influence on the scattering of $m>0$ modes. 
By the symmetry $\{\omega,m\}\to\{-\omega,-m\}$ of the dispersion relation, they will have a similarly important role for scattering of $m<0$ modes.

Firstly, for $\Lambda<\Lambda_c$, the value of $C$ relative to $C_0$ determines when there is a light-ring on the upper branch. 
For $C<C_0$, the light-ring is real and positive; for $C=C_0$, $r_\lr=0$ and for $C>C_0$, $r_\lr$ is complex, i.e. there is no extremum on $\omega^+$.
The absence of a light-ring has very interesting consequences for scattering.
In particular, it means that strong rotation suppresses the propagation of the $+$ and $-$ modes in the vortex core, forcing them to tunnel all the way down to $r=0$.

\begin{figure} 
\centering
\includegraphics[width=\linewidth]{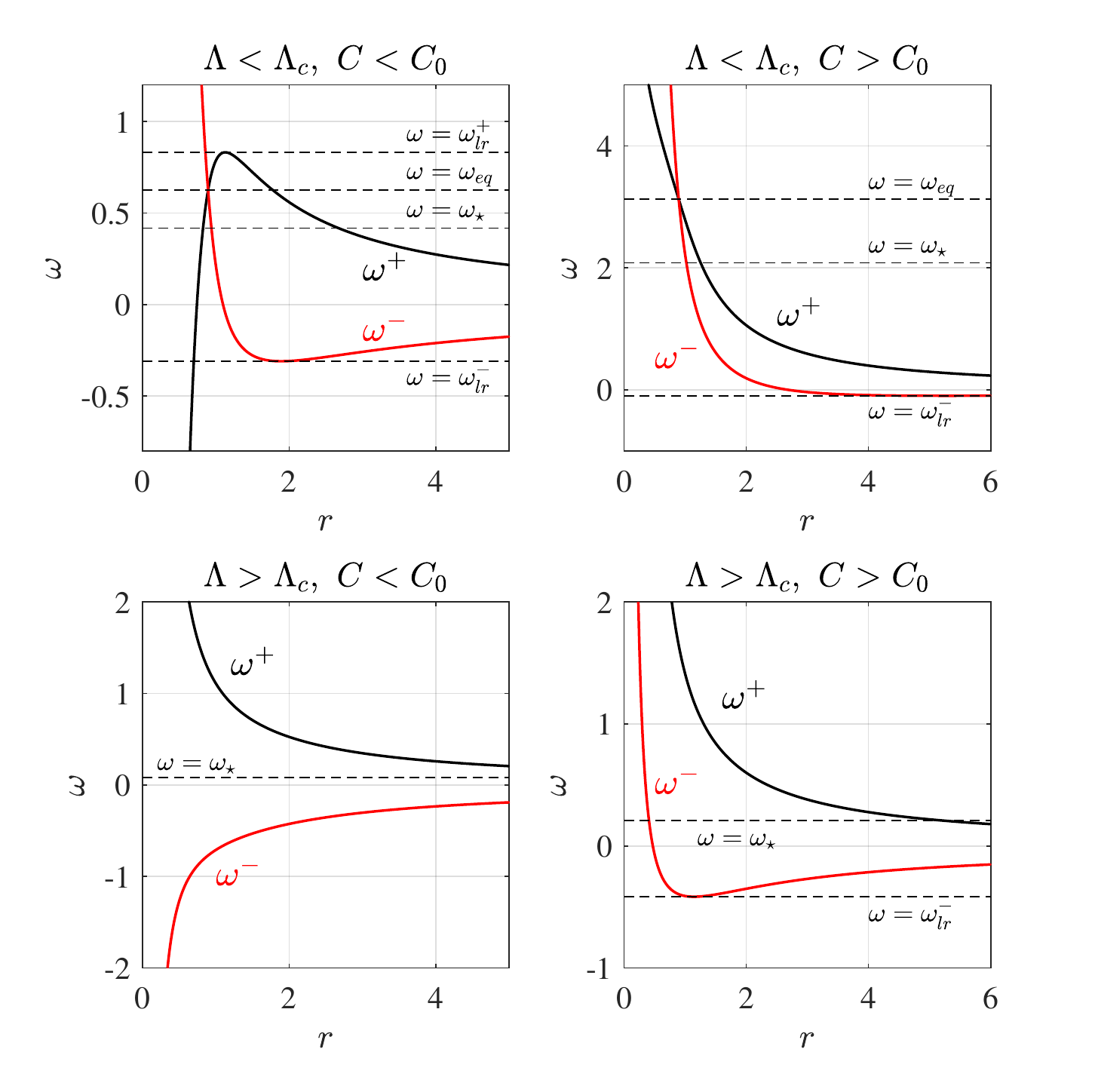}
\caption{The value of $\omega$ at the extrema of $\omega_D^\pm$ for four different cases.
The intersection of a line of $\omega=\cst$ with these curves gives the location of the turning points on the upper and lower branches of the dispersion relation.
Hence one can read these plots to explain why the different scattering types occur in different frequency ranges in Fig.~\ref{fig:paramspace_disp}.
The $\omega^\pm$ are shown here for $m>0$.
However, due to the symmetry $\{\omega,m\}\to\{-\omega,-m\}$ of the dispersion relation, one can simply invert the plots through the horizontal axis (and exchange the $+$ and $-$ labels) to understand the $m<0$ side of the parameter space.
Note that in the top left panel, as $C$ is increased $\omega_\lr^+$ eventually passes inside the $\omega^-$ curve, which corresponds to passing the point where $\omega_\lr=\omega_\eq$ in the top left panel of Fig.~\ref{fig:paramspace_disp}.
The specific parameters $[m,C,\Lambda]$ used to produce these plots were $[1,0.5,0.01]$ (top left), $[1,0.5,2.5]$ (top right), $[1,0.2,0.5]$ (bottom left) and $[1,0.5,0.5]$ (bottom right).} \label{fig:extrema_disp}
\end{figure}

Now consider $\Lambda>\Lambda_c$.
In this case, there is no light-ring on the upper branch of the dispersion relation for any value of $C$.
The value of $C_0$ now determines whether there is an extremum on $\omega^-$; in particular, it is absent for $C<C_0$.
Although we are not interested in the light-ring frequency on the lower branch (since only the $\omega>0$ modes are considered) it's absence is important since it implies that the $+$ and $-$ modes can no longer tunnel to the lower branch of the dispersion relation.
The reason for this is that at large $r$, the $m^2/r^2$ term in $\omega_D^\pm$ will initially cause the two branches to separate.
However, if $\partial_r\omega_D^-$ is nowhere satisfied, the lower branch will continue moving towards lower values of $\omega$ approaching $r=0$, thereby making it impossible for positive frequency modes to reach this branch.
Hence, superradiance (which relies on tunnelling to the $\omega_D^-$ branch) is impossible in this regime.
However, above $C_0$ superradiance will still occur.
The implication of this is that above $\Lambda_c$, each $m>0$ mode has a minimum possible rotation below which it cannot superradiate.
Note that all of the properties discussed above can be deduced using the example $\omega^\pm$ curves given in Fig.~\ref{fig:extrema_disp}.

\subsubsection{Threshold frequency}

As seen in the non-dispersive case, the onset of superradiance is signalled by the tunnelling of the $+$ and $-$ modes from $\omega_D^+$ at large $r$ to $\omega_D^-$ at small $r$.
Let's start by assuming that there exists a threshold frequency $\omega_\star$ that governs when this occurs.
For $\Lambda=0$, it was simple to see what this frequency should be since we only had to look for when two modes appeared on the lower branch of the dispersion relation outside of the horizon.
In the present case, this criterion is not sufficient, since it is now possible for propagating modes to exist on the upper and lower branches of the dispersion relation simultaneously (as demonstrated by Fig.~\ref{fig:disprel_disp}).
Hence, in the dispersive case, one must make sure that it is indeed the $+$ and $-$ modes which tunnel to $\omega_D^-$ and not the $t$ and $b$ modes.

\begin{figure} 
\centering
\includegraphics[width=\linewidth]{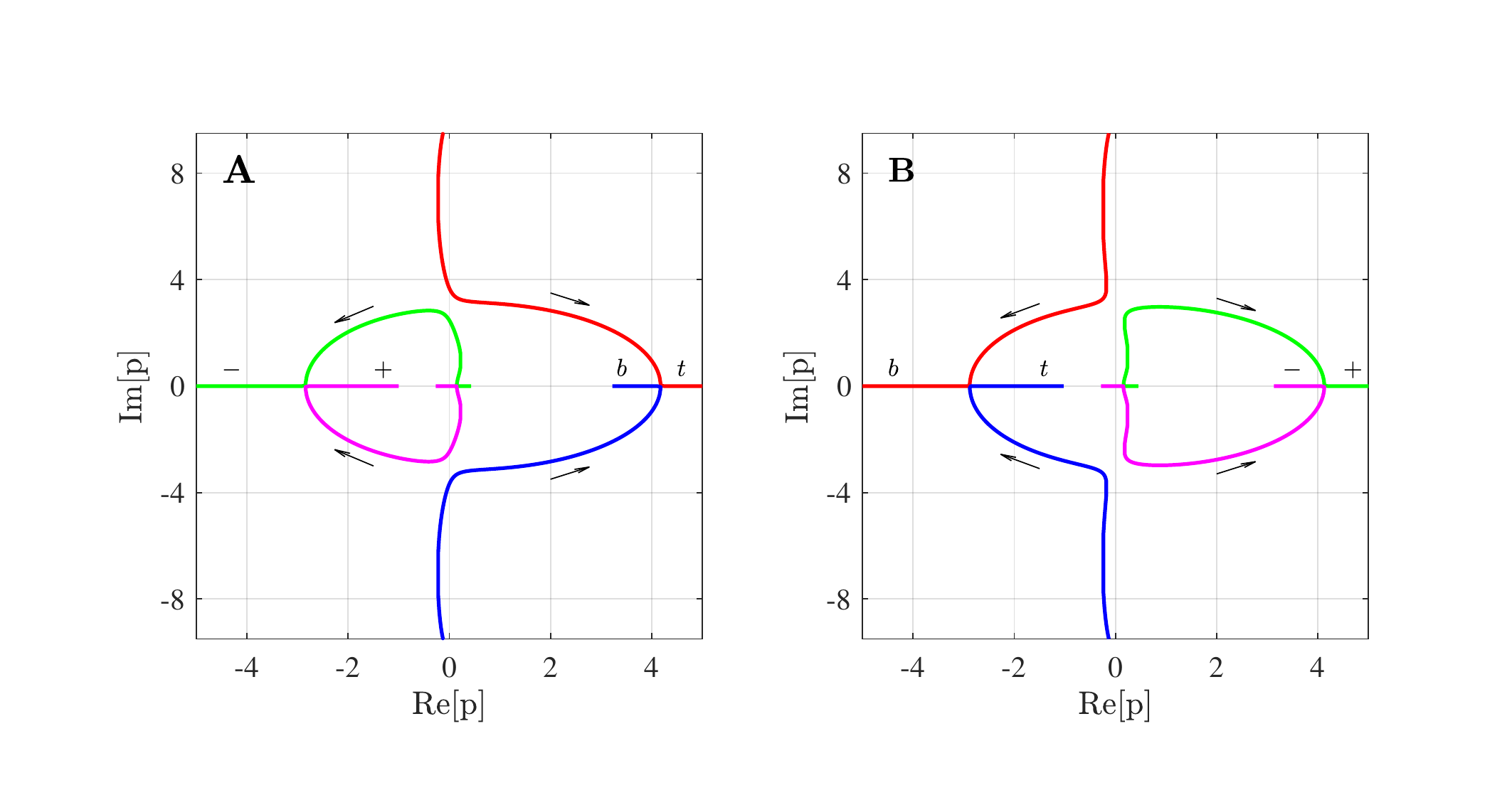}
\caption{The trajectory of the four $p$ solutions to the dispersion relation through the complex $p$-plane, for $\omega<\omega_\star$ (panel A) and $\omega>\omega_\star$ (panel B).
The curves are parametrised by $r$, which decreases in the direction of the arrows.
Observe that as $\omega_\star$ is crossed from below, the red and the green paths (and by symmetry the blue and pink paths) meet at a saddle point in the complex plane. 
As this happens, there is a discontinuous change in the character of the scattering as the evanescent modes at infinity disconnect from the positive frequency branch and reconnect with the negative frequency branch in the vortex core.
The specific parameters are $m=1$, $C=0.5$, $\Lambda=0.01$, with $\omega=0.41$ in A and $\omega=0.425$ in B.
These correspond to scattering Types $\mathrm{IX}^+$ and $\mathrm{VIII}^+_\mathrm{b}$ (defined later) for panels A and B respectively.
} \label{fig:complex_traj}
\end{figure}

To find a necessary condition for this, consider the following argument.
Using only the dispersion relation, one can find an example where $p^\pm$ tunnel to $\omega_D^-$ and one where they do not simply by testing different values of $\omega,m,C,\Lambda$.
The trajectories of the four $p^\s(r)$ through the complex $p$-plane in these examples would look like those shown in Fig.~\ref{fig:complex_traj}.
The difference between these two cases is that $p^\pm$ and $p^{t,b}$ bounce off each other in opposite directions in $p$-plane as $r$ is varied.
This deflection is centred on a saddle point of $\mathcal{H}$ in the complex plane and in the limit that $\omega=\omega_\star$, the modes undergo a head on collision at the point $p_\star$ (and also at the point given by it's complex conjugate).
The task at hand then is to find the expression for these points.

To do this, let us write $p=x+iy$ and $\mathcal{H}(p) = U+iW$, so that,
\begin{equation}
\begin{split}
U = & \ \frac{1}{2}\Big[\Lambda(x^4-6x^2y^2+y^4) + \left(1+2\Lambda m^2/r^2 - 1/r^2\right)(x^2-y^2) \\
& \qquad \qquad \qquad - 2\tilde{\omega}x/r -\tilde{\omega}^2 + \left(1+\Lambda m^2/r^2\right)m^2/r^2\Big], \\
W = & \ -\tilde{\omega}y/r + \left(1+2\Lambda m^2/r^2 - 1/r^2\right)xy + 2\Lambda xy(x^2-y^2).
\end{split}
\end{equation}
Using the fact that $p_\star$ is a saddle point, one has the condition that all $x,y$ derivatives of $U,W$ must vanish, but since $\mathcal{H}(p)$ is holomorphic, the Cauchy-Riemann relations are satisfied and two of these conditions give redundant information.
When the saddle point is a solution to the dispersion relation, $U$ and $W$ also vanish.
Solving these four conditions yields,
\begin{equation}
r_\star = (1+2\Lambda^\frac{1}{2}m)^\frac{1}{2}, \quad \omega_\star = mC/r_\star^2, \quad p_\star = i(m^2+(r_\star^2-1)/2\Lambda)^\frac{1}{2}/r_\star.
\end{equation} 
For small $\Lambda$, the non-dispersive behaviour of the threshold frequency is recovered at low $m$ whereas at high $m$, $\omega_\star$ tends to a constant value of $C/2\Lambda^\frac{1}{2}$.

\subsubsection{Extremum equality}

The equality of the $\omega$ value of the extrema of $\omega_D^\pm$ defines another important frequency $\omega_\eq$.
Using the curves $\omega^\pm$, the condition for this becomes $\omega_\eq=\omega^+=\omega^-$ which, along with the turning point criteria, gives,
\begin{equation}
r_\eq = (1-2\Lambda^\frac{1}{2}m)^\frac{1}{2}, \qquad \omega_\eq = mC/r_\eq^2.
\end{equation}
The importance of this is that it determines the relative location of the turning points on the upper and lower branches of the dispersion relation;
for $\omega>\omega_\eq$, the turning point on the upper branch occurs at larger $r$ than the one on the lower branch and vice versa for $\omega<\omega_\eq$.
This introduces some sub-classification criteria for two of the scattering categories defined in the next section.

\subsubsection{Classification}

\begin{figure} 
\centering
\includegraphics[width=\linewidth]{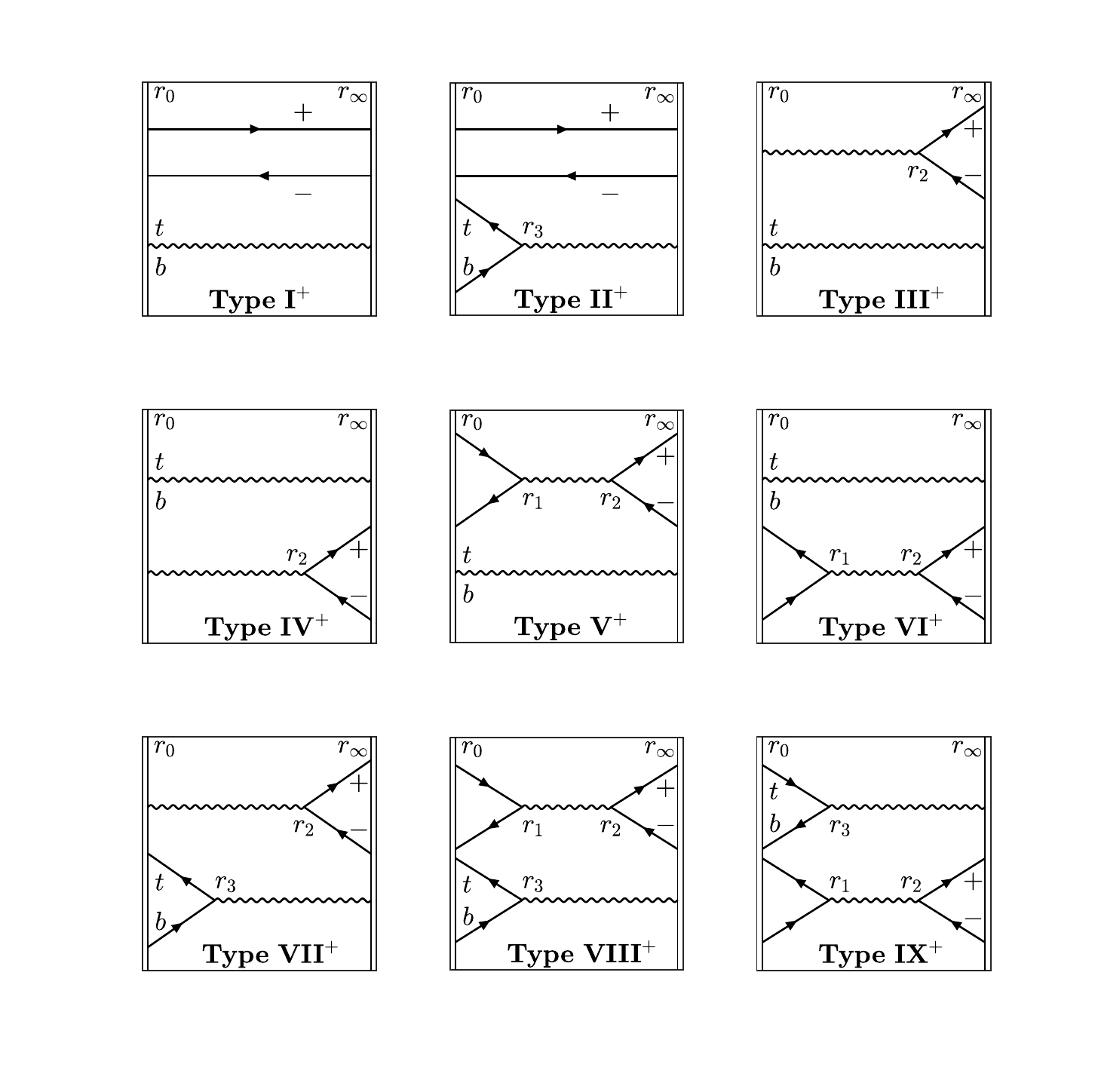}
\caption{An illustration of the trajectories of the four WKB modes through the $(r,p)$ phase space, with $r$ on the horizontal axis and $p$ on the vertical.
As shown in Fig.~\ref{fig:paramspace_disp}, the different types occur in different regions of the $(C,\omega,\Lambda)$ parameter space.
In particular, superradiance occurs in Type $\mathrm{VI}^+$ and $\mathrm{IX}^+$. 
} \label{fig:feynman2}
\end{figure}

In Fig.~\ref{fig:feynman2}, a schematic illustration is provided of the real part of $p^\s(r)$ through the $(r,p)$ phase space for the four different modes.
Again the use of these diagrams is in identifying how many turning points there are (and which modes they involve), as well as the asymptotic behaviour of the modes.
The turning points are labelled as follows:
$r_1$ and $r_2$ are the inner and outer turning points for the $+,-$ modes, where $r_3$ is the only allowed turning point for the $t,b$ modes.

\begin{figure} 
\centering
\includegraphics[width=\linewidth]{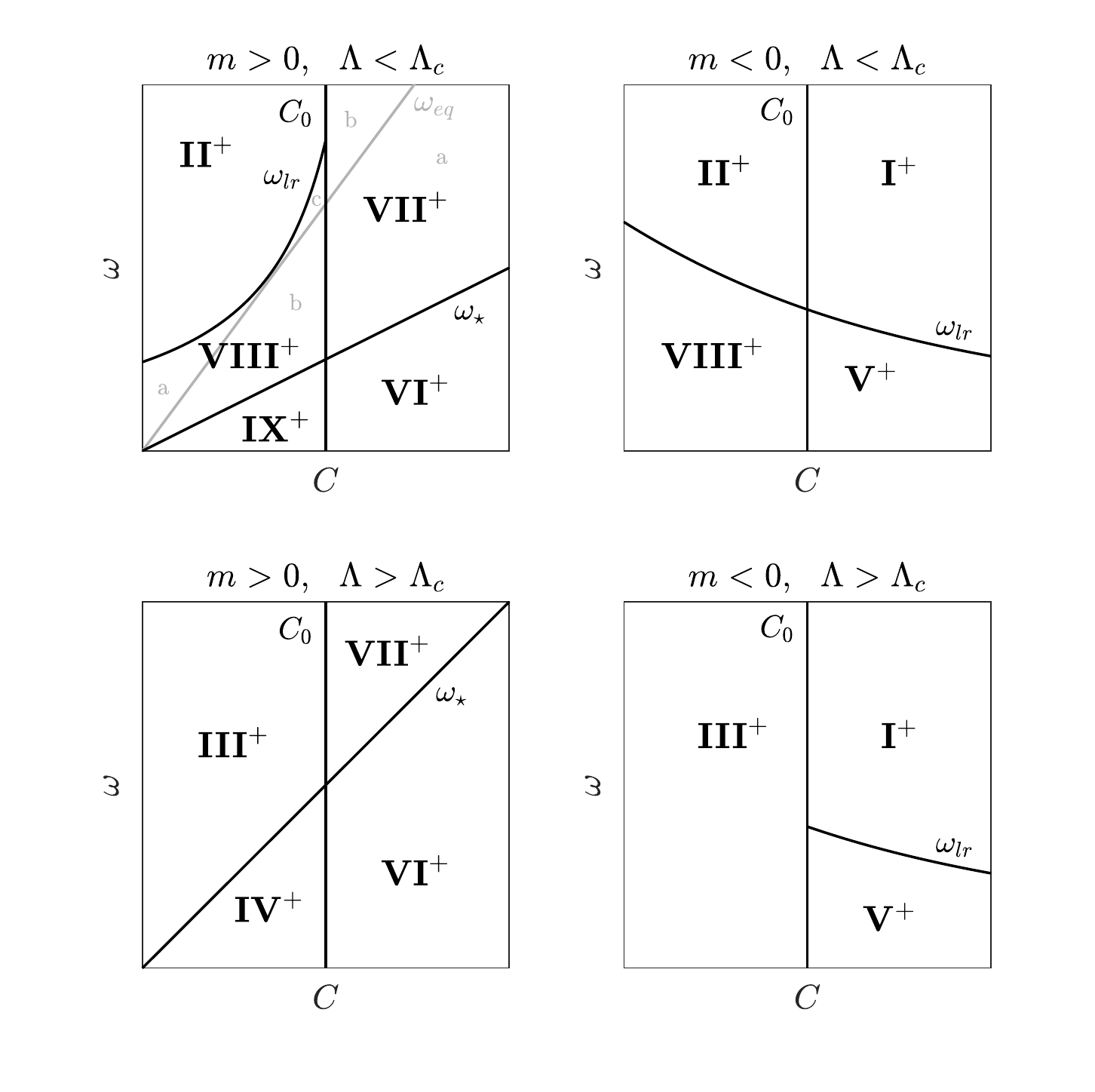}
\caption{Different types of scattering are shown to occur in different regions of the $(C,\omega)$ parameter space (note that the bottom left corner is the point $(0,0)$).
There is a distinct change in the structure of the parameter space depending on the size of $\Lambda$ relative to the critical value $\Lambda_c$. 
The reason for the different scattering types can be understood by considering Fig.~\ref{fig:extrema_disp} and observing where the turning points are for a particular $\omega$.
} \label{fig:paramspace_disp}
\end{figure}

These diagrams represent the type of scattering that occurs in different regions of the $(C,\omega)$ parameter space, which is depicted in Fig.~\ref{fig:paramspace_disp}.
The parameter space has a distinctively different structure depending on the sign of $m$, as well as the size of $\Lambda$ relative to $\Lambda_c$.
To understand this structure, it is again useful to plot the $\omega$ value of the extrema $\omega_D^\pm$,
i.e. $\omega^\pm(r)=\omega_D^\pm(r,\pm p_\ex(r))$.
The turning points (and their relative locations) can then be deduced by looking for the intersections of these curves with a line $\omega=\cst$.
There are four distinct cases for the shapes of $\omega^\pm$, which are illustrated in Fig.~\ref{fig:extrema_disp}.

A couple of noteworthy points concerning the diagrams in Fig.~\ref{fig:feynman2}.
Firstly, there is a sub-classification for the Type $\mathrm{VII}^+$ and $\mathrm{VIII}^+$ diagrams depending on the relative locations of $r_{1,2,3}$.
Specifically,
\begin{itemize}
\item Type $\mathrm{VII}^+_\mathrm{a}$: $r_3<r_2$,
\item Type $\mathrm{VII}^+_\mathrm{b}$: $r_2<r_3$,
\item Type $\mathrm{VIII}^+_\mathrm{a}$: $r_3<r_1<r_2$,
\item Type $\mathrm{VIII}^+_\mathrm{b}$: $r_1<r_3<r_2$,
\item Type $\mathrm{VIII}^+_\mathrm{c}$: $r_1<r_2<r_3$.
\end{itemize}
Note, however, that this does not alter the computation of the scattering coefficients since the relative location of $r_3$ to $r_{1,2}$ does not play a role in this.

Secondly, the diagrams give the impression that the modes with larger $\mathrm{Re}[p]$ appear at the top and smaller $\mathrm{Re}[p]$ are toward the bottom.
This ordering is accurate whilst ever the modes are real, but not necessarily when they are complex.
For instance, in Type $\mathrm{VIII}^+$ scattering it happens that $\mathrm{Re}[p^+]>\mathrm{Re}[p^{t,b}]>\mathrm{Re}[p^-]$ as $r\to\infty$, but $\mathrm{Re}[p^+]>\mathrm{Re}[p^-]>\mathrm{Re}[p^{t,b}]$ as $r\to 0$ (compare Fig.~\ref{fig:mode_trajs} to the Type $\mathrm{VIII}^+$ diagram for a clear example of this) which is allowed since the $t$ and $b$ modes are complex as they cross the real modes (really they move around them in the complex $p$-plane).
This is not a problem since the use of the diagrams is to determine the mode asymptotics and identify the turning points.
In fact, this overlapping of real parts is purposefully not shown in Fig.~\ref{fig:feynman2} so that it is clear which modes are interacting (i.e. share common turning points) and which are not.

With this reshuffling of real parts allowed for complex modes, the diagrams for Types $\mathrm{III}^+$ and $\mathrm{IV}^+$ technically represent the same class of scattering.
The reason that they have been left as separate cases is the following: in Type $\mathrm{III}^+$, the $+$ and $-$ modes try to tunnel to $\omega_D^+$ approaching the origin, but do not make it since the branch recedes from the modes due to dispersion. In Type $\mathrm{IV}^+$ they try instead to tunnel to $\omega_D^-$.
In this sense, Type $\mathrm{IV}^+$ represents failed superradiance, since the negative energy mode which would otherwise propagate into the vortex core is forbidden from doing so due to the strength of dispersion. 

\subsection{Scattering coefficients}

Following the procedure outlined in Section~\ref{sec:scat_coefs}, I will now be interested in writing down the relations between the different scattering coefficients to show the existence of superradiance.
There are four independent solutions to the radial equation of motion, $j=1,2,3,4$, which can be defined by their asymptotics.
For each diagram in Fig.~\ref{fig:feynman2}, one could in priciple write down an asymptotic formula to define the modes as in \eqref{asymp}.
However, this would be a tedious process and most of the solutions written down would contain no more information than that which is readily apparent from looking at the diagrams.
In the following, I will therefore make some simplifying observations to avoid having to write down each the solutions separately.
It will then become apparent that the important scattering coefficients, and the relations between them, are exactly the same as in the non-dispersive case.

Firstly, each scattering type falls into one of four classes with different mode asymptotics, which can be easily identified by looking at how the mode trajectories approach $r_0$ and $r_\infty$ in the diagrams in Fig.~\ref{fig:feynman2}.
These classes are outlined in Table~\ref{tab:asymp}.
Note that the diagrams under asymptotic classes A and B can be covered by a single formula in the same way that \eqref{asymp} was used to represent all three diagrams in the non-dispersive case.

\begin{table}[ht]
\centering 
\begin{tabular}{c c c c} 
\hline \hline 
Class & $r_0~\{+,-,t,b\}$ & $r_\infty~\{+,-,t,b\}$ & Types \\ [0.5ex] 
\hline 
A & $\{\mathbb{R,R,C,C}\}$ & $\{\mathbb{R,R,C,C}\}$ & $\mathrm{I}^+,\mathrm{V}^+,\mathrm{VI}^+$ \\
B & $\{\mathbb{R,R,R,R}\}$ & $\{\mathbb{R,R,C,C}\}$ & $\mathrm{II}^+,\mathrm{VIII}^+,\mathrm{IX}^+$ \\
C & $\{\mathbb{C,C,C,C}\}$ & $\{\mathbb{R,R,C,C}\}$ & $\mathrm{III}^+,\mathrm{IV}^+$ \\
D & $\{\mathbb{C,C,R,R}\}$ & $\{\mathbb{R,R,C,C}\}$ & $\mathrm{VII}^+$ \\ [1ex] 
\hline 
\end{tabular}
\caption{The diagrams in Fig.~\ref{fig:feynman2} are classified according to whether each of the WKB modes is asymptotically propagating or evanescent.
$\mathbb{R}$ indicates that a particular mode is propagating, i.e. $p^\s\in\mathbb{R}$, where as $\mathbb{C}$ is used for evanescent modes, i.e. $p^\s\in\mathbb{C}$.} 
\label{tab:asymp}
\end{table}

Next, it is easy to see that in each digram, the $+,-$ modes do not interact with the $t,b$ modes (this contrasts what happens when dispersion is subluminal and all four modes can interact \cite{patrick2020superradiance,patrick2020quasinormal}).
Therefore, two of the $R_j$ will be on the $+,-$ part of the diagram (say $R_{1,2}$) and the remaining two will be in the $t,b$ part.
There are two possibilities for the interaction of the $t,b$ modes: either they are evanescent everywhere and do not interact, in which case $R_{3,4}$ are the pure WKB $t,b$ modes over the whole region and the amplitudes are unrelated; or the propagating $t,b$ modes in the core are completely reflected, in which cases the $R_{3,4}$ are analogous the $\mathrm{Ai}$ and $\mathrm{Bi}$ solving Airy's equation with the amplitudes related by a phase shift.
In both of these cases, no energy is carried by the $t,b$ modes across the system, hence, they will be of no further interest from here on.

Lastly we have the interaction of $+,-$ modes.
In asymptotics classes C and D, there is a complete reflection of these modes at large $r$ and the two independent solutions are again analogous to the Airy functions $\mathrm{Ai}$ and $\mathrm{Bi}$. 
In asymptotics classes A and B, the $+,-$ mode asymptotics are identical to those in the non-dispersive case given in \eqref{asymp}, this time with the upper sign taken for Types $\mathrm{I}^+$, $\mathrm{II}^+$, $\mathrm{V}^+$ and $\mathrm{VIII}^+$ and the lower sign for Types $\mathrm{VI}^+$ and $\mathrm{IX}^+$.
The relations between the different amplitudes are again given by \eqref{scatterfull} and following the analysis just below \eqref{asymp}, one finds that superradiance occurs for Types $\mathrm{VI}^+$ and $\mathrm{IX}^+$.
In fact, there is a simpler way to see this directly from Fig.~\ref{fig:feynman2}, since any diagram which contains an interaction of the form,
\begin{equation}
\includegraphics[scale=0.3]{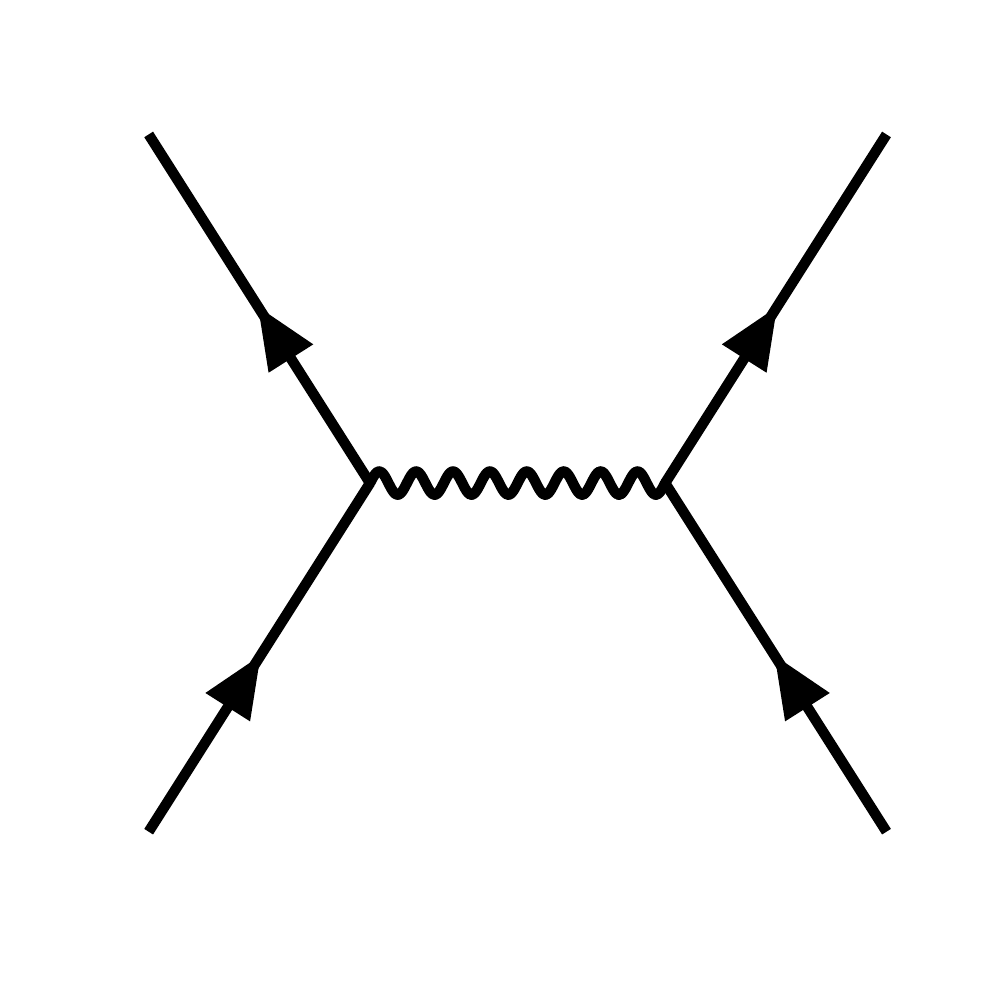}
\end{equation}
will be superradiant.
The reason for this is that this relative orientation of the four arrows is the smoking gun for tunnelling between the upper and lower branches of the dispersion relation.
As discussed previously, a mode which propagates into the vortex core on $\omega_D^-$ carries in a negative energy, which by energy conservation means that the escaping mode must be amplified.

Finally, it is simple matter to deduce an expression for the reflection coefficient from the diagrams.
For Types $\mathrm{I}^+$ and $\mathrm{II}^+$, $\mathcal{R}$ vanishes whereas for Types $\mathrm{III}^+$, $\mathrm{IV}^+$ and $\mathrm{VII}^+$ it is simply $|\mathcal{R}|=1$.
In the remaining cases, $|\mathcal{R}|$ is obtained from \eqref{LocalScatter} as,
\begin{equation} \label{refl2}
|\mathcal{R}| = \left(\frac{1-f_{12}^2/4}{1+f_{12}^2/4}\right)^{\pm 1},
\end{equation}
with the upper sign taken for Types $\mathrm{V}^+$ and $\mathrm{VIII}^+$ and the lower sign for Types $\mathrm{VI}^+$ and $\mathrm{IX}^+$.
Hence, the form of the reflection coefficient is identical to the non-dispersive case, and the only difference is in the location of the turning points $r_{1,2}$ and the integral of the phase between these points.

\begin{figure} 
\centering
\includegraphics[width=\linewidth]{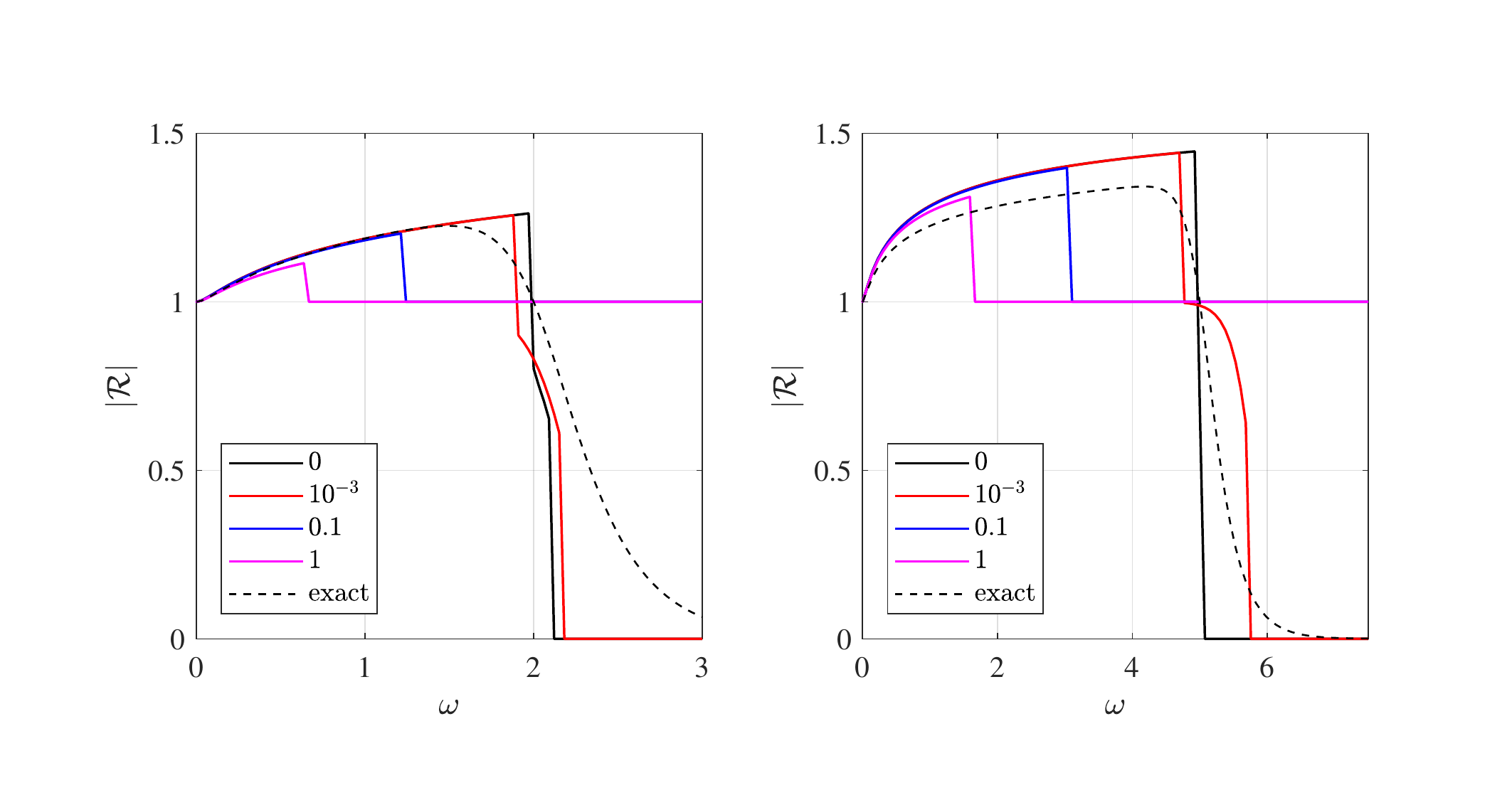}
\caption{Examples of the reflection coefficient for the $m=1$ mode for $C=2$ (left) and $C=5$ (right).
The different curves represent different values of $\Lambda$.
The main impact of dispersion on superradiance is to reduce the bandwidth of superradiant modes; the amount of amplification is affected much less.
An exact numerical computation of $|\mathcal{R}|$ in the non-dispersive case (following Appendix A of \cite{patrick2020superradiance}) is shown as a dashed curve for comparison.
Note that the discontinuous changes are an artefact of only considering scattering due to turning points, and in reality will be smoothed over by other (sub-dominant) sources of scattering.
} \label{fig:refls}
\end{figure}

The behaviour of $|\mathcal{R}|$ can be easily read off from Fig.~\ref{fig:paramspace_disp} using the knowledge gained from the diagrams in Fig.~\ref{fig:feynman2}.
For the positive $m$'s (which are the ones which can superradiate) there are only three distinct behaviours.
For $\Lambda<\Lambda_c$ and $C<C_0$, the modes are amplified at low $\omega$ and absorbed at high $\omega$, much like the non-dispersive case.
For $C>C_0$ for any $\Lambda$, the modes are amplified at small $\omega$ and completely reflected for high $\omega$.
Finally, for $\Lambda>\Lambda_c$ and $C<C_0$, the modes are completely reflected for all $\omega$.
Some examples of this behaviour are given in Fig.~\ref{fig:refls}.
It is clear that for superradiant modes, dispersion needs to be strong before any siginificant change arises to the amount of amplication.
By contrast, changes to the bandwidth of superradiant modes become evident even for small $\Lambda$.

Note that the WKB approximation has a tendency to over-estimate $|\mathcal{R}|$ when compared with the exact non-dispersive result.
In fact, this is to be expected since the true potential appearing in the non-dispersive radial equation is a modification of \eqref{eff_potential} by $m^2/r^2\to (m^2-1/4+5 v_r^2/4)/r^2$ \cite{dolan2012resonances}.
When this is used in the approximate formula for $|\mathcal{R}|$ (which amounts to applying the WKB approximation in the radial direction only), the exponent appearing in $f_{12}$ becomes larger, which has the effect of reducing deviations of the reflection coefficient away from unity.
As promised, these differences quickly become very small in the limit of large $m$.

\subsection{Comments} \label{sec:posL_comments}

Although the scattering classification for the dispersive waves was much more involved that for non-dispersive ones, there are only a few differences when it comes to understanding superradiance in the two cases.
These have already been alluded to earlier, but I re-emphasize them here for clarity.

When dispersion is weak, i.e. $\Lambda<\Lambda_c$, superradiance proceeds for all positive frequencies below the threshold frequency $\omega_\star$ for any value of the rotation parameter.
However, unlike the non-dispersive case, the superradiant bandwidth does not continue to grow with $m$, and instead levels off at a constant value of $C/2\Lambda^\frac{1}{2}$ in the limit that $m\to\infty$.
One would therefore expect the energy extracted by a given $m$-mode to be less with dispersion than without it. 

When dispersion is strong, i.e. $\Lambda>\Lambda_c$, there is another important difference.
In this regime, the $\omega\in[0,\omega_\star]$ modes can only superradiate above a critical rotation parameter given by $C_0(m)$.
Assuming that superradiance will cause the vortex to spin down (this is revisited in Section \ref{sec:discuss}), there will be a point along time evolution of $C$ where each $m$ stops superradiating.
Since, $C_0$ is an increasing function of $m$ in this regime, only the small $m$ modes superradiate at low $C$.
In particular, if $\Lambda>1/4$ then the $m=1$ will not be able to superradiate, which implies that \textit{there will be no superradiant modes in the system.}
The consequence of this is that for $\Lambda<1/4$, the system will eventually spin-down to zero rotation since there is always an angular momentum channel into which the vortex can dissipate.
However, for $\Lambda>1/4$ there is a minimum allowed rotation given by $C_\mathrm{min}=(4\Lambda-1)/4\Lambda^\frac{1}{2}$, and once this value is reached, superradiance shuts off completely.

In the next section, I will show that the vortex will shed energy and angular momentum spontaneously as a result of superradiance acting upon it's vacuum fluctuations.


\section{Spontaneous emission} \label{sec:spont}

Due to quantum fluctuations in the system, there will be a steady flux of energy and angular momentum out of the vortex, even when the fields $\phi$ and $\eta$ are in their vacuum states.
To compute this quantum emission, I will work in the fully dispersive case.
This is principally done so that both dispersive and non-dispersive cases can be handled simultaneously (the non-dispersive result is found as the $\Lambda\to 0$ limit of the dispersive one).
Physically, however, the dispersive case is the one relevant to real experiments, where the effects of dispersion are never completely negligible.

\subsection{Quantisation}

The quantisation procedure given here follows that of \cite{unruh1974second}.
The conjugate momentum to $\phi$ is obtained from \eqref{action} as,
\begin{equation}
\pi = \frac{\partial\mathcal{L}}{\partial\dot{\phi}} = -\eta .
\end{equation}
Canonical quantisation then proceeds by replacing complex conjugation by Hermitian conjugation, promoting the fields $\phi$ and $\eta$ to operators and imposing the equal time commutation relations,
\begin{equation}
\begin{split}
[\hat{\phi}(x),\hat{\phi}(y)] = & \ [\hat{\pi}(x),\hat{\pi}(y)] = 0, \\
[\hat{\phi}(x),\hat{\pi}(y)] = & \ i\delta^{(2)}(x-y).
\end{split}
\end{equation}
In \eqref{comp_pos_neg}, the amplitudes $\alpha^*_\lambda$ and $\alpha_\lambda$ and are replaced by creation and annihilation operators, $\hat{a}^\dagger_\lambda$ and $\hat{a}_\lambda$ respectively, which obey,
\begin{equation}
[\hat{a}_{\lambda_1},\hat{a}^\dagger_{\lambda_2}] = \delta_{\lambda_1\lambda_2}.
\end{equation}
The normal mode expansion of the fields then reads,
\begin{equation} \label{NM_expans}
\hat{\phi} = \sum_\lambda (\hat{a}_\lambda \varphi_\lambda + \hat{a}^\dagger_\lambda \varphi^*_\lambda), \qquad \hat{\eta} = \sum_\lambda (\hat{a}_\lambda n_\lambda + \hat{a}^\dagger_\lambda n^*_\lambda).
\end{equation}
The vacuum is defined as the state which is annihilated by all $\hat{a}_\lambda$, i.e. $\hat{a}_\lambda|0\rangle = 0$.

The goal is to compute the vacuum expectation value (VEV) of the energy and angular momentum current far away from the vortex.
To do this, an expression is needed for the energy and angular momentum current operators.
Starting from the expressions in \eqref{e_curr} and \eqref{l_curr}, define the following operators,
\begin{equation} \label{quantumEcurr_general}
\widehat{\textfrak{S}}_E^r = \tfrac{1}{2}\big(\tfrac{1}{2}v_r\{\hat{\phi},\partial_t\hat{\eta}\}-\tfrac{1}{2}v_r\{\hat{\eta},\partial_t\hat{\phi}\} - \{\partial_r\hat{\phi},\partial_t\hat{\phi}\} - \Lambda\{\partial_r\hat{\eta},\partial_t\hat{\eta}\}\big),
\end{equation}
and,
\begin{equation} \label{quantumLcurr_general}
\widehat{\textfrak{S}}_L^r = \tfrac{1}{2}\big(\tfrac{1}{2}v_r\{\hat{\phi},\partial_\theta\hat{\eta}\}-\tfrac{1}{2}v_r\{\hat{\eta},\partial_\theta\hat{\phi}\} - \{\partial_r\hat{\phi},\partial_\theta\hat{\phi}\} - \Lambda\{\partial_r\hat{\eta},\partial_\theta\hat{\eta}\}\big),
\end{equation}
where $\{~,~\}$ is the anti-commutator.
This ordering of terms is chosen to make the operator symmetric in the fields whilst still recovering \eqref{e_curr} and \eqref{l_curr} in the classical limit.

Inserting the expansions in \eqref{NM_expans}, the VEVs of the different anti-commutators appearing in Eq.~\eqref{quantumEcurr_general} are,
\begin{equation} \label{VEVs}
\begin{split}
\langle 0|\{\hat{\phi},\partial_t\hat{\eta}\} |0\rangle = & \ -i\sum_\lambda\nolimits \omega\left[n_\lambda\varphi_\lambda^* - \varphi_\lambda n^*_\lambda\right], \\
\langle 0|\{\hat{\eta},\partial_t\hat{\phi}\} |0\rangle = & \ -i\sum_\lambda\nolimits \omega\left[\varphi_\lambda n^*_\lambda - n_\lambda\varphi_\lambda^*\right], \\
\langle 0|\{\partial_r\hat{\phi},\partial_t\hat{\phi}\}|0\rangle = & \ -i\sum_\lambda\nolimits \omega\left[\varphi_\lambda \partial_r\varphi^*_\lambda - (\partial_r\varphi_\lambda)\varphi_\lambda^*\right], \\
\langle 0|\{\partial_r\hat{\eta},\partial_t\hat{\eta}\}|0\rangle = & \ -i\sum_\lambda\nolimits \omega\left[n_\lambda \partial_r n^*_\lambda - (\partial_rn_\lambda)n_\lambda^*\right].
\end{split}
\end{equation}
The anti-commutators in Eq.~\eqref{quantumLcurr_general} are given by similar relations except there, the $\omega$ sitting inside of the sum is replaced by $m$.
The flux of energy and angular momentum out of the vortex are the quantities of interest, hence one must integrate the VEV of \eqref{quantumEcurr_general} and \eqref{quantumLcurr_general} around a ring far away from the origin.
Noticing that the four terms in \eqref{VEVs} are simply those appearing in \eqref{Wronsk}, one finds,
\begin{equation} 
\int^{2\pi}_0 d\theta ~r \langle 0 | \widehat{\textfrak{S}}_E^r | 0 \rangle = 2\pi \sum_\lambda \omega W[\varphi_\lambda,\varphi_\lambda],
\end{equation}
and similarly for the angular momentum current, where the integral has been evaluated by noticing that $W[\varphi_\lambda,\varphi_\lambda]$ is independent of $\theta$.

To perform the normal mode sum, one can make a few simplifying observations.
Firstly, notice that $W[\tilde{\varphi}_j,\tilde{\varphi}_j]$ vanishes for solutions which are everywhere evanescent and those which are completely reflected.
This means that none of the diagrams in asymptotics classes C or D will contribute.
For the same reason, the $R_{3,4}$ solutions in asymptotics classes A and B will also not contribute.
Thus, only the $R_{1,2}$ solutions, given in \eqref{asymp}, need to be taken into consideration.
Taking the expressions for $W[\tilde{\varphi}_1,\tilde{\varphi}_1]$ and $W[\tilde{\varphi}_2,\tilde{\varphi}_2]$ evaluated at $r_\infty$ directly from \eqref{scatterfull}, one arrives at,
\begin{equation}
\begin{split}
2\pi \sum_j W[\tilde{\varphi}_j,\tilde{\varphi}_j] = & \ \frac{1}{2\pi}\left[\frac{q_\infty^- + q_\infty^+|\mathcal{R}|^2}{|q_\infty^-|} + \frac{q^+_\infty|\mathcal{U}|^2}{|q^\pm_0|}\right], \\
= & \ \frac{|\mathcal{R}|^2-1}{2\pi} \left[ 1+\mathrm{sgn}(q^\pm_0/q^-_\infty)\right],
\end{split}
\end{equation}
where in the second line, I have used $q^+_\infty q^-_\infty|\mathcal{U}|^2=q^\pm_0(q_\infty^- + q_\infty^+|\mathcal{R}|^2)$, which is obtained by combining the three relations in \eqref{scatterfull}, and also the fact that the reflection coefficient is evaluated just outside $r_2$ so $q_\infty^+=-q_\infty^->0$.
Recalling that the upper sign should be taken for scattering Types $\mathrm{I}^0$, $\mathrm{II}^0$, $\mathrm{I}^+$, $\mathrm{II}^+$, $\mathrm{V}^+$ and $\mathrm{VIII}^+$ and the lower sign for Types $\mathrm{III}^0$, $\mathrm{VI}^+$ and $\mathrm{IX}^+$, one finds $\mathrm{sgn}(q^\pm_0/q^-_\infty)=\mp 1$.
Thus, the term in the square brackets vanishes for all non-superradiant scattering scenarios.

Finally, equating this energy flux with an energy loss from the system (and the same for the angular momentum flux) one finds,
\begin{equation} \label{rates}
\begin{pmatrix}
\dot{E} \\ \dot{L}
\end{pmatrix} = -\frac{1}{\pi}\sum_m\int_{SR} d\omega ~ \begin{pmatrix}
\omega \\ m
\end{pmatrix}
\left(|\mathcal{R}|^2-1\right),
\end{equation}
in agreement with \cite{unruh1974second}.
Here $SR$ indicates that the $\omega$ integral is performed only over superradiant frequencies, and the expression for $|\mathcal{R}|$ is given by that in \eqref{refl2} with the lower sign.

\begin{figure} 
\centering
\includegraphics[width=\linewidth]{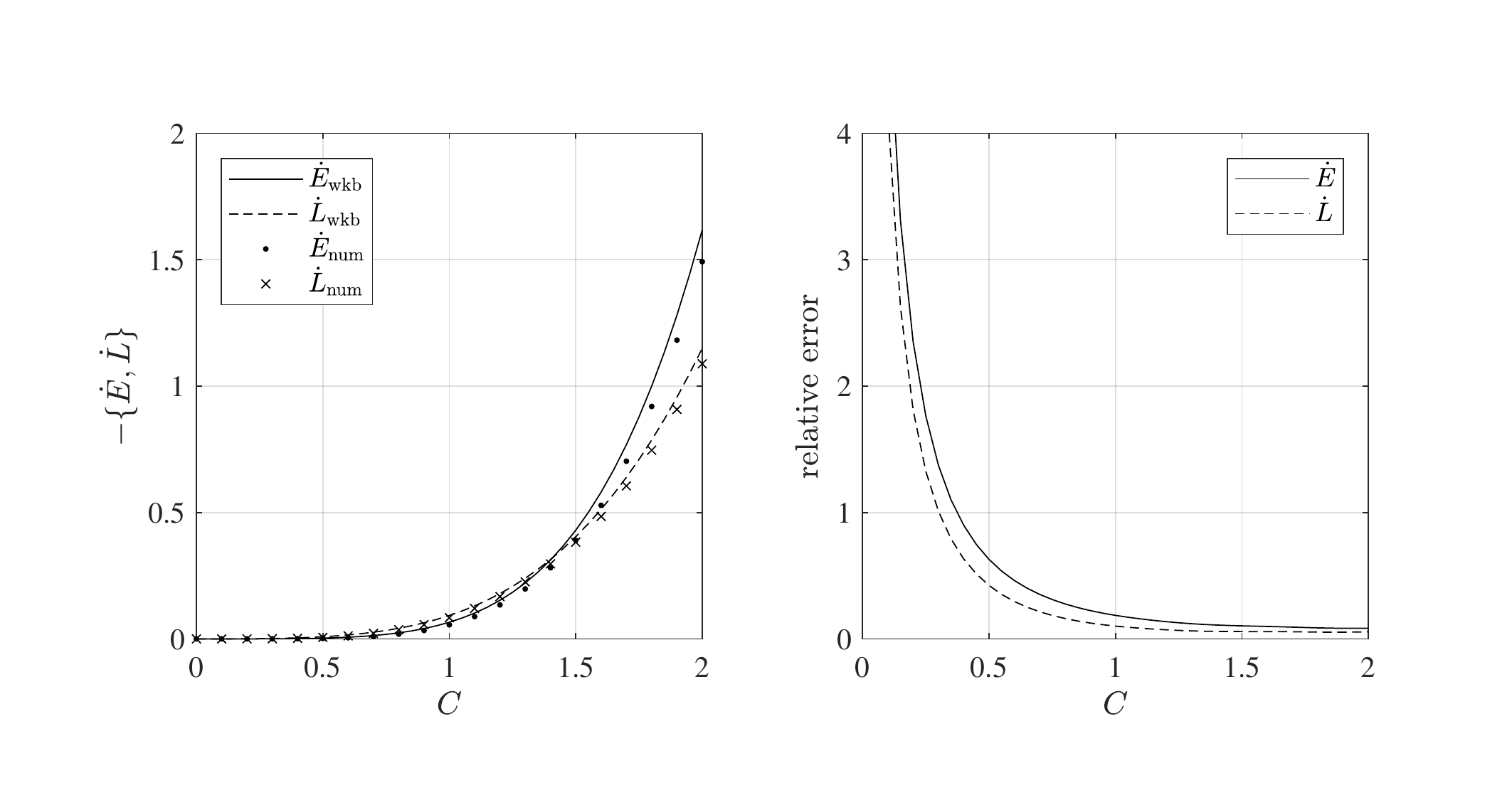}
\caption{The values of $\dot{E}$ and $\dot{L}$ as a function of $C$ using the WKB approximation (solid and dashed lines respectively) and an exact result obtained from numerical simulation of the equations of motion (dots and crosses respectively).
The relative error between the WKB and exact result is a decreasing function of $C$, indicating that the approximation improves for large $C$.} \label{fig:ELrates}
\end{figure}

\subsection{Non-dispersive results}

\begin{figure} 
\centering
\includegraphics[width=.5\linewidth]{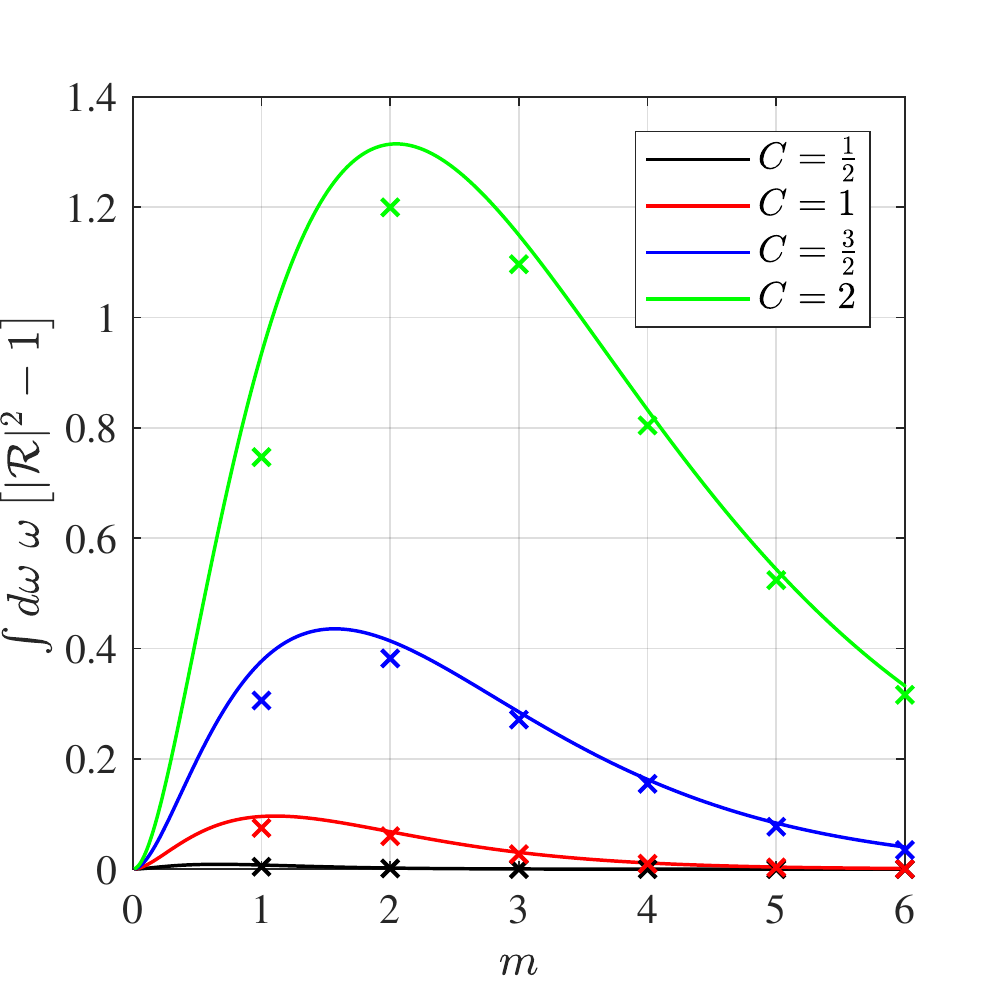}
\caption{The integral appearing in $\dot{E}$ for different values of $m$ and $C$.
The solid line respresents the WKB reflection coefficients, whereas the crosses are from numerical simulation.
Although \eqref{rates} is a sum over integer values of $m$, the WKB result is extended to non-integer values here to clearly indicate the location of the peak, which appears at roughly $m\sim C$.
Since the WKB result becomes closer to the exact one for higher $m$, the agreement between $\dot{E}_\mathrm{num}$ and $\dot{E}_\mathrm{wkb}$ will improve for large $C$ where the higher $m$ modes dominate the sum.
A similar plot can be obtained for the components of $\dot{L}$.
} \label{fig:Ecomps}
\end{figure}

To evaluate $\dot{E}$ and $\dot{L}$, the reflection coefficients are computed using the WKB formula in Eq.~\eqref{refl2}, as well as from exact simulation of the wave equation (see e.g. Appendix A of \cite{patrick2020superradiance}).
Fig.~\ref{fig:ELrates} demonstrates that the rates of energy and angular momentum loss are dramatically increased for larger $C$ values (in dimensionful units, this means for large $C/D$ ratios).
It is also shown that the agreement between the exact and WKB results increases with $C$.
The reason for this is dominant mode in the sum in \eqref{rates} is the $m\sim|C/D|$ mode as shown in Fig.~\ref{fig:Ecomps}. Since the WKB approximation improves for large $m$, the values of $\dot{E}$ and $\dot{L}$ will become more accurate for large $C$.
This contrasts the black hole case where the lowest angular momentum mode always dominates the sum \cite{unruh1974second}.
However, this is to be expected since in a Kerr black hole $a/M<1$, where $a$ and $M$ are the rotation and mass parameters of the spacetime.
The difference with a fluid mechanical vortex is that the ratio $C/D$ is (in principle) not bound from above.
In the large $C$ limit, the rates are approximately,
\begin{equation}
\dot{E} \approx -0.05~C^5, \qquad \dot{L} \approx -0.07~C^4,
\end{equation}
where the coefficients and exponents have been obtained from a numerical fit over the range $C\in[0,100]$.

\subsection{Dispersive results}

\begin{figure} 
\centering
\includegraphics[width=\linewidth]{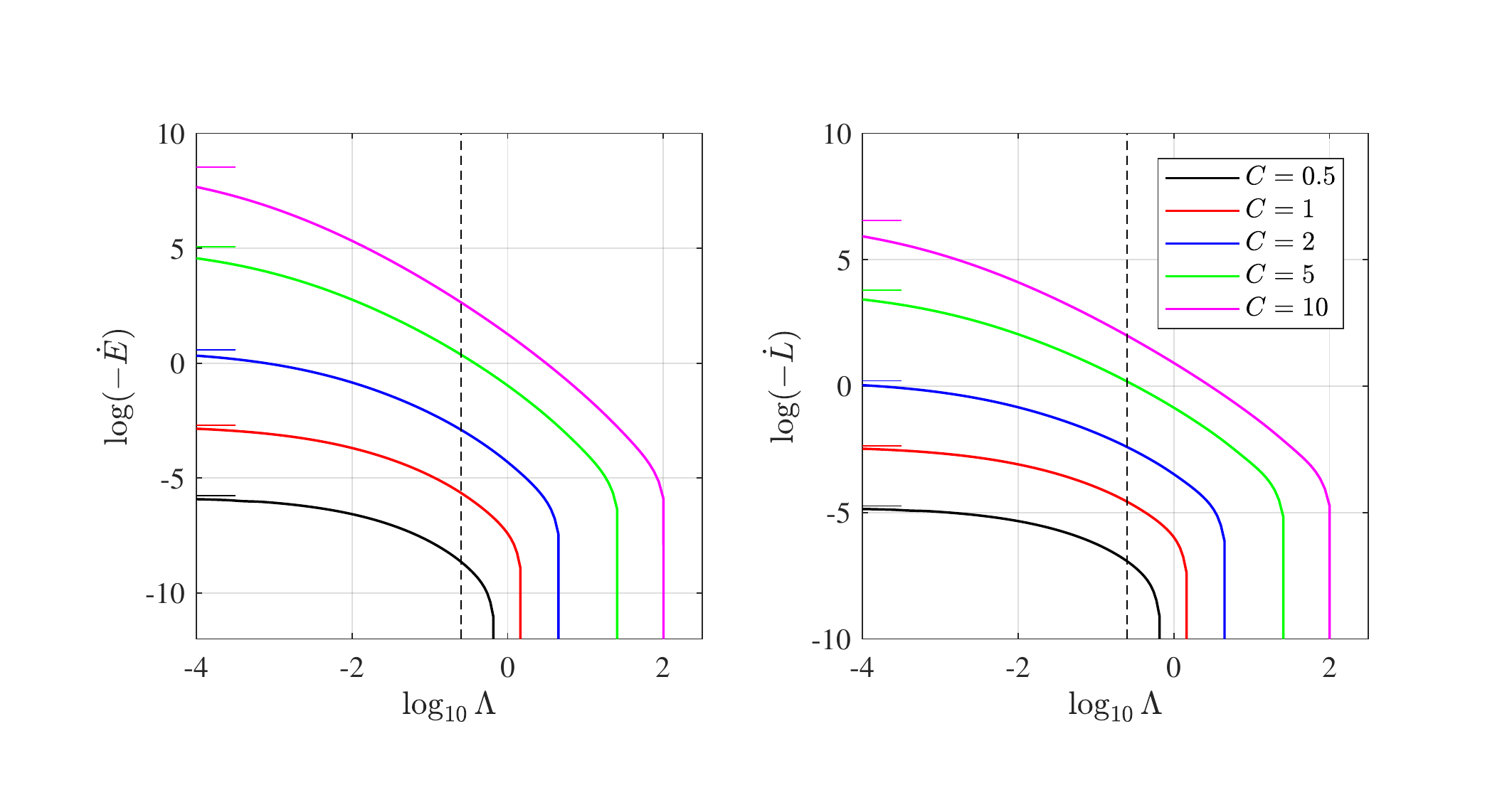}
\caption{Dispersion decreases the rate at which superradiance extracts energy and angular momentum from the vortex. 
Non-dispersive rates  are indicated as horizontal lines on the left of the plots for comparison.
The critical $\Lambda$ value of $\Lambda=1/4$ is shown as a dashed black line.
} \label{fig:Ldep_rates}
\end{figure}

When $\Lambda>0$, the dependence of $\dot{E}$ and $\dot{L}$ on $C$ is qualitatively similar to the non-dispersive case.
Fig.~\ref{fig:Ldep_rates} demonstrates that as the parameter $\Lambda$ increases, the values of $\dot{E}$ and $\dot{L}$ decrease.
As explained at the end of Section~\ref{sec:posL_comments}, there are two principle reasons for this; 1) the superradiant bandwidth becomes narrower as $\Lambda$ increases and 2) modes with $m$ larger than $m_\mathrm{max}$ (obtained by inverting the expression for $C_0$ in \eqref{critical_vals}) will not superradiate.
For fixed $C$, $m_\mathrm{max}$ decreases with increasing $\Lambda$ until eventually there are no superradiant modes left in the system, and $\dot{E}$ and $\dot{L}$ go to zero.
Inverting this statement, for $\Lambda$ fixed (and larger than $1/4$) there is a value $C_\mathrm{min}$ below which superradiance will not occur for any $m$.
$C_\mathrm{min}$ goes to zero at $\Lambda=1/4$, which means that for $\Lambda<1/4$, superradiance will occur at least for some $m$ modes for all $C$.

\section{Discussion} \label{sec:discuss}

\subsection{Effect of quantised $\ell$}

In contrast to previous sections, from here on the dimensions on various quantities will be restored.
In analogy with the circulation parameter defined in \eqref{circ}, let the drain parameter be given by,
\begin{equation}
D = \hbar d/M,
\end{equation}
where (unlike $\ell$) $d$ is not constrained to be an integer.
There are two reasons to write $D$ this way.
The first is that the ratio $C/D=\ell/d$, which permits an easy comparison of the drain strength to the winding number.
The second is that $d=r_h/\xi$ gives the ratio of the ``would be'' horizon (i.e. $D/c$) to the healing length $\xi=\hbar/\sqrt{M\mu}$, which characterises the scale over which the  condensate heals back to it's bulk value around boundaries. 
In particular $\xi$ gives a characteristic length on which the density varies in the vortex core.

The dimensionless dispersion parameter discussed throughout this work is now really $\Lambda/D^2=1/4d^2$.
Therefore, the relative strength of dispersion is controlled by how fast the system is draining.
Importantly, it was noted above that superradiance shuts off at a critical $C$ value when $\Lambda/D^2>1/4$.
In terms of $\ell$ and $d$, this really means that when the system is weakly draining, i.e. $d<1$, superradiance can only spin down the system whilst $\ell>\ell_\mathrm{min} = (1-d^2)/2$.
However, since $l$ can only take on integer values, the minimum rotation for a spinning vortex will be $\ell=1$, which is always greater than $\ell_\mathrm{min}$. Thus, the vortex will still be able to superradiate whilst ever it is spinning, irrespective of the value of $d$.

Although the rates in \eqref{rates} will force the system to evolve, it remains unclear exactly how this evolution might occur.
From $\mathcal{S}_\mathrm{GPE}$, the conserved current for rotations in $\theta$ gives the angular momentum of the system,
\begin{equation}
L = \hbar \ell N.
\end{equation}
This quantity must decrease as a result of $\dot{L}$ in \eqref{rates}.
There are two ways this may occur; either by a reduction in the number of atoms in the condensate or a reduction in $\ell$.
If it is the former, then the rotation $C$ remains fixed during the evolution, whereas if it is the latter then the density decreases.
To determine which of these is the dominant effect, one would need to solve the full backreaction equations.
Furthermore, since $\ell$ is quantised, it can only presumably only decrease in discrete lumps, but this may itself induced problems since it would require an instantaneous global change in the phase of the condensate.
One possibility is that evolution may proceed through the emission of a quantised vortex.
The precise detail of this evolution certainly warrants further investigation.


\subsection{Gravity waves}

It was argued above that the suppression of superradiance due to strong dispersion (weak drain) would not be observable in a BEC due to the quantised nature of circulation.
Might there be another system where this suppression is observable?

The equations of motion in \eqref{wave_eqn} are equivalent to those governing capillary-gravity waves in the shallow water regime $hk\ll 1$, with $h$ the height of the fluid.
In that case, the wave speed is $c=\sqrt{gh}$, with $g$ now the acceleration due to gravity, and the dispersive parameter is $\Lambda=\sigma h/\rho_f$, with $\sigma$ the surface tension of the fluid and $\rho_f$ it's density. 
In the experiment of \cite{torres2017rotational}, $D$ and $h$ were of the order of $10^{-3}$ and $10^{-2}$ respectively.
Taking the values of $\sigma$ and $\rho_f$ for water, the dimensionless dispersive parameter is $\Lambda/D^2\sim\mathcal{O}(1)$.
At this strength of dispersion, the $m=1,2$ modes would cease to superradiate around $C/D\sim\mathcal{O}(1)$.
However, since $C/D$ was closer to 15 for that experiment, one would still expect to see superradiance for these modes.
This is consistent with observation.

Note that the present estimates are for shallow water waves, whereas the experiments in \cite{torres2017rotational} were performed closer to the deep water regime.
It was shown in \cite{patrick2020superradiance} that superradiance can still occur in deep water, and a next natural step would be to add in the capillary modification to the dispersion relation.
Since this makes the dispersion superluminal at high $k$, one would expect some of the features of scattering in the present system to carry over.
In particular, one might expect superradiance to be suppressed for low circulation.
Importantly, since the fluid is classical, there is no longer a minimum value for $C$, raising the possibility of observing this suppression of superradiance in a water tank experiment akin to that of \cite{torres2017rotational}.

\section{Conclusion}

In this work, a general framework for studying superradiance in dispersive systems (which follows on from \cite{patrick2020superradiance}) has been presented.
This framework was then applied to study how superradiance occurs under Bogoliubov type dispersion.
Through a detailed analysis of the possible scattering outcomes, it was found that the superradiance condition in \eqref{sr_bh} gets modified to,
\begin{equation} \label{sr_mod}
\omega<m\Omega_h^\mathrm{rot}\left(1+m\xi/r_h\right)^{-1},
\end{equation}
where $r_h$ and $\Omega_h^\mathrm{rot}$ are defined as in the non-dispersive case, although $r_h$ is no longer a true horizon.
This condition essentially determines when the in-going mode ceases to tunnel back to the $\Omega>0$ branch of the dispersion relation at small $r$, and instead attempts to tunnel to the $\Omega<0$ branch.
However, \eqref{sr_mod} is no guarantee that the mode will have reached this branch by $r=0$, which means as a consequence that (in this particular regime) incident waves are completely reflected by the vortex and superradiance suppressed.
In turned out that this suppression of superradiance was outside of the parameter range for superfluids owing to the effects of quantised circulation and an analogous classical system, namely shallow water capillary-gravity waves, was instead suggested as an alternative where this suppression may be observable.

It was then argued that the main influence of dispersion on the superradiance spectrum would be to reduce the bandwidth in $\omega$ over which superradiance occurs, whereas the amount of amplification for a given superradiant mode would not change significantly.
The physical reason for this is that the modes involved in the scattering are the same ones present in the non-dispersive case, which live on the part of the dispersion relation where the effects of $\Lambda\neq 0$ are smallest.
In particular, the location of the turning points (which determine size of the reflection coefficient) are only slightly modified.
A further difference with the non-dispersive system is that just above the superradiance threshold, the combination of large rotations and dispersion meant that incident $m>0$ modes would not be absorbed by the vortex and instead completely reflected.

Finally, the spontaneous emission of the vortex due to it's superradiant vacuum fluctuations was studied.
As expected from classical considerations, it was found that the rate of energy and angular momentum loss is decreased by dispersion, since there are less $\omega$ and $m$ modes to superradiate.
Another interesting finding was that (in the non-dispersive case where all $m>0$ superradiate) the mode which extracts most from the system is the $m\sim|C/D|$ mode.
This highlights an interesting property of superradiance which is not present in black hole physics, where the equivalent ratio ($a/M$) is always less than unity of a consequence of cosmic censorship.

For simplicity, I have only treated the constant density approximation here, due to the possibility of obtaining concise formulae for the various important parameters (in particular $\omega_\star$).
Since $\rho(r)$ varies over a (minimum) scale of $\sim\xi$, the constant density approximation is applicable roughly when $r_h\gg\xi$, or equivalently $\Lambda/D^2\ll 1/4$.
This has implications for the regions of parameter space in Fig.~\ref{fig:paramspace_disp} where the scattering coefficients provided herein are a good approximation.
Despite this, the effects of varying $\rho$ could easily be incorporated into the formalism, simply by inserting a coefficient of $\rho(r)/\rho(\infty)$ in front of the $k^2$ term in \eqref{branches}, and the predictions improved.


\ack
Many thanks to Cisco Gooding, August Geelmuyden, Sebastian Erne and Silke Weinfurtner for discussions involving various aspects of this project.
Financial support was provided by Silke Weinfurtner's Leverhulme Research Leadership Award (RL-2019-020).

\appendix
\section{} \label{app:norm}
In this appendix, the normalisation of incident WKB modes (which defines the different independent solutions, e.g. \eqref{asymp}, of the radial equation of motion) is performed explicitly.
The inner product \eqref{norm1} evaluated for two WKB modes is,
\begin{equation}
\begin{split}
(\varphi_{\lambda_1},\varphi_{\lambda_2}) = & \ \frac{1}{2}\int d^2\mathbf{x}~ \frac{\mathcal{N}_{\lambda_1}\mathcal{N}^*_{\lambda_2}}{r}\left|\frac{f_{\lambda_1}f_{\lambda_2}}{\mathcal{H}'_{\lambda_1}\mathcal{H}'_{\lambda_2}}\right|^\frac{1}{2} \left(\frac{\Omega_{\lambda_1}}{f_{\lambda_1}}+\frac{\Omega_{\lambda_2}}{f_{\lambda_2}}\right)e^{i\int(p_{\lambda_1}-p_{\lambda_2})dr+i(m_1-m_2)-i(\omega_1-\omega_2)t}, \\
= & \ \tfrac{1}{2}(2\pi)\delta_{m_1m_2}e^{-i(\omega_1-\omega_2)t}\mathcal{N}_{\lambda_1}\mathcal{N}^*_{\lambda_2} \int dr \left|\frac{f_{\lambda_1}f_{\lambda_2}}{\mathcal{H}'_{\lambda_1}\mathcal{H}'_{\lambda_2}}\right|^\frac{1}{2} \left(\frac{\Omega_{\lambda_1}}{f_{\lambda_1}}+\frac{\Omega_{\lambda_2}}{f_{\lambda_2}}\right)e^{i\int(p_{\lambda_1}-p_{\lambda_2})dr},
\end{split}
\end{equation}
where the $\mathcal{N}_\lambda$ are constants to be found from normalisation, and to get to the second line the $\theta$ integral has been evaluated.
At this point, it would be possible to evaluate the $r$ integral if the exponent were a difference of constants multiplied by $r$ and the prefactor under the integral were also constant. 
This is indeed the case when the modes are normalised in the non-dispersive case at spatial infinity.
However, it is not the case (in general) when the modes are dispersive, or are being normalised in a region where the background flow is non-vanishing.

To proceed, recognise that in the WKB approximation, $p_{\lambda_1}$ and $p_{\lambda_2}$ are assumed large.
Hence if $p_{\lambda_1}\neq p_{\lambda_2}$, the phase term will oscillate rapidly leading to cancellation in the integral.
The dominant contribution comes from near $p_{\lambda_1}=p_{\lambda_2}$.
This cannot be satisfied unless one works with the same type of WKB mode, but this defines the specific $j$ solution of the radial equation of motion.
Hence the expression above picks up a factor of $\delta_{j_1j_2}$.
Next, expand the momentum,
\begin{equation}
p(\omega_1) = p(\omega_2) + \partial_\omega p(\omega_2) (\omega_1-\omega_2),
\end{equation}
and similarly for $\Omega_1$, $f_1$ and $\mathcal{H}'_1$.
Note, I have dropped the subscript $\lambda_{1,2}$ since it is now assumed we are working with the same $m,j$-mode and hence, the different quantities only differ due to their dependence on $\omega$.
At leading order in $(\omega_1-\omega_2)$, the integral becomes,
\begin{equation}
2\int dr \left| \frac{f(\omega_2)}{\mathcal{H}'(\omega_2)}\right|\frac{\Omega(\omega_2)}{f(\omega_2)}e^{i(\omega_1-\omega_2)\int\partial_\omega p(\omega_2)dr}.
\end{equation}
By noting $f>0$ for propagating modes, the $f$ terms cancel.
Next, define a new coordinate $X=\int\partial_\omega p(\omega_2)dr$, which satisfies $dr = \vec{\mathbf{e}}_r\cdot\bm{v}_g~dX$.
The integral now reads,
\begin{equation}
\mathrm{sgn}(\Omega)~2\int dX e^{i(\omega_1-\omega_2)X} = \mathrm{sgn}(\Omega)~4\pi\delta(\omega_1-\omega_2),
\end{equation}
where to get to the left hand side, I have used $\mathcal{H}'=\vec{\mathbf{e}}_r\cdot\bm{v}_g\Omega$.
Plugging this back into the expression above, one obtains,
\begin{equation}
(\varphi_{\lambda_1},\varphi_{\lambda_2}) = \mathrm{sgn}(\Omega)(2\pi)^2 \mathcal{N}_{\lambda_1}\mathcal{N}^*_{\lambda_2} \delta_{m_1m_2}\delta_{j_1j_2}\delta(\omega_1-\omega_2).
\end{equation}
Thus, for these modes to be normalised, the amplitude must be,
\begin{equation}
\mathcal{N}_\lambda = (2\pi)^{-1},
\end{equation}
and the normalisation condition can be written in the compact notation,
\begin{equation}
(\varphi_{\lambda_1},\varphi_{\lambda_2}) = \mathrm{sgn}(\Omega)\delta_{\lambda_1\lambda_2}.
\end{equation}

\section*{References}
\bibliographystyle{iopart-num}
\bibliography{superradiance_quartic_final.bbl}
\end{document}